\documentclass[twocolumn,prb,superscriptaddress]{revtex4-1}
\usepackage{graphicx,bm,amsmath,amssymb,xcolor,soul,tikz,pgfplots,array}
\usepackage{scrextend}
\usepackage{epstopdf}

\let\hide\iffalse

\newcommand{\nk}[1]{{\color{black}#1}}

\def\F{Fr\"ohlich}
\def\ef{E_{\rm F}}
\def\W0{\Omega_0}
\def\w0{\hbar\omega_0}
\def\kf{k_{\rm F}}
\newcolumntype{P}[1]{>{\centering\arraybackslash}p{#1}}

\def\oden{Oden Institute for Computational Engineering and Sciences, 
The University of Texas at Austin, 201 E. 24$^{th}$ Street, 
Austin, TX 78712, USA}
\def\physics{Department of Physics, The University of Texas at Austin, 
Austin, TX 78712, USA}
\def\oxford{Department of Materials, University of Oxford, Parks Road, 
Oxford OX1 3PH, United Kingdom}
\def\vienna{Fakult{\"a}t f{\"u}r Physik, Universit{\"a}t Wien, 
Boltzmanngasse 5, 1090 Vienna, Austria}

\begin{document}

\title{Many-body Green's function approaches to the doped \F\ solid:\\[4pt] 
Exact solutions and anomalous mass enhancement}

\author{Nikolaus Kandolf}
\affiliation{\oden} \affiliation{\physics} \affiliation{\oxford}
\author{Carla Verdi}
\affiliation{\vienna}
\author{Feliciano Giustino}
\email{fgiustino@oden.utexas.edu}
\affiliation{\oden} \affiliation{\physics}

\date{\today}

\begin{abstract}
In polar semiconductors and insulators, the \F\ interaction between electrons and long-wavelength 
longitudinal optical phonons induces a many-body renormalization of the carrier effective masses
and the appearence of characteristic phonon sidebands in the spectral function, commonly dubbed 
`polaron satellites'. The simplest model that captures these effects is the \F\ model, whereby
electrons in a parabolic band interact with a dispersionless longitudinal optical phonon.
The \F\ model has been employed in a number of seminal papers, from early perturbation-theory
approaches to modern diagrammatic Monte Carlo calculations. One limitation of this model 
is that it focuses on undoped systems, thus ignoring carrier screening and Pauli blocking effects 
that are present in real experiments on doped samples. To overcome this limitation, we here extend the 
\F\ model to the case of doped systems, and we provide exact solutions for the electron spectral 
function, mass enhancement, and polaron satellites. We perform the analysis using two approaches, 
namely Dyson's equation with the Fan-Migdal self-energy, and the second-order 
cumulant expansion. We find that these two approaches provide qualitatively different results. 
In particular, the Dyson's approach yields better quasiparticle masses and worse satellites, 
while the cumulant approach provides better satellite structures, at the price of worse quasiparticle 
masses. Both approaches yield an anomalous enhancement of the electron effective mass at
finite doping levels, which in turn leads to a breakdown of the quasiparticle picture in
a significant portion of the phase diagram.
\end{abstract}

\maketitle

\section{Introduction}

The \F\ interaction, that is the coupling between electrons and long-wavelength longitudinal 
optical (LO) phonons in polar semiconductors and insulators, constitutes one of the earliest and
most intensely studied manifestations of electron-phonon physics.\cite{Frohlich1950,Devreese2000,
Devreese2009} On the theory side, \F\ couplings have received considerable attention during 
the past few years, as efficient \textit{ab initio} techniques to describe these processes 
have become available.\cite{Verdi2015,Sjakste2015,Park2020,Brunin2020}
\nk{Meanwhile, a recent report has demonstrated the remarkable 
effectiveness of a generalized \F\, model in the prediction of zero-point 
band-gap renormalization when compared to highly accurate \textsl{ab-initio} 
calculations.\cite{Miglio2020}}
On the experiment side, the \F\ interaction has long been known to play an important role 
in the carrier transport properties of doped semiconductors and oxides~\cite{Ohtomo2004,
Herranz2007} and in their superconducting phases.\cite{Schooley1964,Schooley1965,Lin2014} 
More recently, \F\ couplings have been identified as the origin of intriguing phonon sidebands 
in the photoelectron spectra of many compounds, including SrTiO$_3$ (STO),\cite{Meevasana2011,
Santander-Syro2011,Chen2015, WangZ2016} TiO$_2$,\cite{Moser2013}, EuO,\cite{Riley2018} 
CaMnO$_3$,\cite{Husanu2020} and ZnO.\cite{Yukawa2016} In turn, these sidebands have been 
linked to the observation of superconducting phases in bulk and interfacial systems.\cite{Lee2014,
Lin2014,Cancellieri2016}

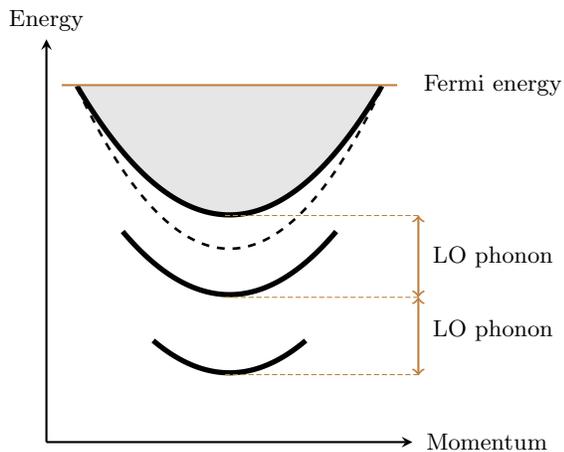
\begin{figure}
  \begin{center}
  \resizebox{0.9\columnwidth}{!}{
  \begin{tikzpicture}
   \begin{axis}[xmin=-1.2,xmax=1.2,ymin=-0.5,ymax=2,
    width=6.5cm,height=7cm,axis line style={line width=0.8pt},
    xtick=\empty, ytick=\empty, minor tick num=0, scaled   
    ticks=false, xticklabel=\empty,yticklabel=\empty,axis x 
    line=bottom,axis y line=left,xlabel={Momentum}, 
    ylabel={Energy},
    every axis x label/.style={at={(current axis.right of 
    origin)},anchor=west,right=10mm,above=-13mm},
    every axis y label/.style={at={(current axis.north 
    west)},above=0mm}]
    \addplot[domain=-1:1,samples=100,black,line width =
    1pt,smooth,no marks, dashed] {0.7+x^2};
    \addplot[domain=-1:1,samples=100,black,line width =
    2pt,smooth,no marks,fill=black!10] {0.91+0.8*x^2};
    \addplot[domain=-0.7:0.7,samples=100,black,line width =
    2pt,smooth,no marks] {0.415+0.8*x^2};
    \addplot[domain=-0.5:0.5,samples=100,black,line width =
    2pt,smooth,no marks] {-0.07+0.8*x^2};
    \addplot[domain=-1.1:1.1,samples=100,brown,line width =
    0.8pt,smooth,no marks] {1.715+0*x};
   \end{axis}
   \draw[<->,color=brown,line width =0.8pt] (5,1.95) -- (5,3.05);
   \draw[<->,color=brown,line width =0.8pt] (5,0.9) -- (5,1.95);
   \draw[-,color=brown,dashed,dash pattern=on 2pt off 1pt] (2.4,0.91) -- (5,0.91);
   \draw[-,color=brown,dashed,dash pattern=on 2pt off 1pt] (2.4,1.95) -- (5,1.95);
   \draw[-,color=brown,dashed,dash pattern=on 2pt off 1pt] (2.4,3.05) -- (5,3.05);
   \node[draw=none] at (6,2.5) {LO phonon};
   \node[draw=none] at (6,1.5) {LO phonon};
   \node[draw=none] at (6,4.8) {Fermi energy};
  \end{tikzpicture}
  }
\end{center}
\caption{Schematic illustration of the many-body renormalization of the conduction
band bottom of a polar semiconductor or insulator by the \F\ interaction.
The illustration refers to a parabolic conduction band minimum, doped with
electrons up to the Fermi level. The dashed line indicates the non-interacting
band structure, the solid lines show the renormalized band minimum as well as
the phonon sidebands (two sidebands for example). The energy separation between
the quasiparticle band and the sidebands is an integer multiple of the LO phonon
	energy.}
\label{fig1}
\end{figure}

The most direct route to investigate the effect of \F\ couplings on electron band structures
is via angle-resolved photoelectron spectroscopy (ARPES). In ARPES experiments, electrons 
are extracted from a sample via laser or synchrotron light, and the energy and momentum
of the electron prior to exiting the sample can be reconstructed by an analyzer. This setup
provides a direct image of the momentum-resolved electron spectral function, i.e. the many-body
electron band structure. Since only occupied electronic states can be probed by ARPES, 
it is necessary to dope electrons into the sample in order to image the band edges.
These electrons interact with all phonons in the sample; however, in polar semiconductors and 
oxides the dominant coupling mechanism is the \F\ interaction with long-wavelength LO phonons, 
because the associated coupling matrix element diverges at long wavelength.\cite{Giustino2017} 
This coupling enhances the carrier effective mass and leads to the appearence of phonon sidebands 
below the conduction band edge, usually called `polaron satellites'.\cite{Moser2013,WangZ2016,
Lee2014,Cancellieri2016} A schematic illustration of these effects is shown in Fig.~\ref{fig1}.

The description of these low-energy structures using \textit{ab initio} many-body methods 
is challenging.\cite{Story2014} In the case of standard metals, where the Fermi energy
$E_{\rm F}$ is much larger than the characteristic phonon energy $\hbar\omega_{\rm ph}$,
$\hbar\omega_{\rm ph}/E_{\rm F} \ll 1$, Migdal's theorem guarantees that the 
interaction is well described by non-crossing electron-phonon self-energy diagrams.\cite{Migdal1958,
Mahan1981,Giustino2017} However, degenerate semiconductors including doped oxides typically
possess a Fermi energy comparable to the characteristic phonon energy, $\hbar\omega_{\rm ph}/E_{\rm F} 
\sim 1$.\cite{BretzSullivan2019} This scenario falls outside of the validity limit of the
Migdal approximation. As a result, calculations on these systems based on the Migdal
approximation suffer from well-documented shortcomings, for example incorrect energetics
of the satellite structures shown in Fig.~\ref{fig1}.\cite{Story2014,Verdi2017,Nery2018,Zhou2019}

One promising strategy to overcome this limitation is provided by the cumulant expansion 
method.\cite{Langreth1970,Hedin1980,Aryasetiawan1996} In its original formulation, this approach 
was introduced to study the coupling of core holes to plasma excitations in metals.\cite{Langreth1970}
The generalization of this approach to valence electrons \cite{Hedin1980} enabled the first 
\textit{ab initio} calculations of plasmon satellites in elemental metals.\cite{Aryasetiawan1996} 
More recently, the cumulant approach has been employed to improve the description of spectral
satellites arising from electron-plasmon interactions in GW calculations.\cite{Lischner2013,Kas2014,
Caruso2015a,Caruso2015b,Caruso2016a,Caruso2016b,Gumhalter2016,Zhou2018b,Tzavala2020}
In the context of electron-phonon physics, the cumulant expansion has successfully been employed 
to calculate phonon sidebands in systems exhibiting \F\ coupling.\cite{Story2014,Verdi2017,Caruso2018,
Riley2018,Nery2018,Zhou2019,Antonius2020,Zhou2020} 

Despite much progress on the front of \textit{ab initio} calculations, we still lack a simple 
analytical model that captures the essential features of \F\ interactions in doped systems, 
and that can be used as a reference benchmark for validating \textit{ab initio} implementations.
This gap is particularly critical as \textit{ab initio} calculations of \F\ couplings require
extremely dense Brillouin zone grids, and are therefore computationally very demanding. As a result, it is
difficult to systematically explore the parameter space and extract general trends.

The most popular model employed to investigate electronic couplings to polar LO phonons is the \F\ 
model.\cite{Frohlich1950} This model consists of an electron in a parabolic electron band
coupled to a dispersionless LO phonon.\cite{Frohlich1950,LeeT1952,Roseler1968,Smondyrev1986,
Selyugin1989,Feynman1954} It forms the basis for a number of seminal papers on 
electron-phonon interactions and polarons,\cite{Devreese2000,Devreese2009,Mahan1981} and is
routinely used for testing advanced many-body techniques such as the diagrammatic Monte Carlo 
method.\cite{Prokofev1998,Mishchenko2000} However, this model describes a single electron
coupled to a phonon bath, therefore it does not include the effects of band filling
(shown schematically in Fig.~\ref{fig1}), and the associated screening of the polar interaction
by free carriers. Without including free carriers, the \F\ model cannot reproduce the correct
energetic ordering of bands and satellites as shown in Fig.~\ref{fig1} and observed in
experiments; instead, the \F\ model incorrectly yields satellites \textit{above} the conduction band bottom.\cite{Mishchenko2000,Nery2018}

In this work, we go beyond the original \F\ model by deriving analytic expressions for the
electron self-energy and spectral function for electron-LO phonon interactions in the presence 
of free carriers. This model constitutes an idealization of \F\ interactions in many degenerate
semiconductors and doped oxides that have been investigated via photoelectron spectroscopy.\cite{Chen2015, 
WangZ2016,Moser2013,Riley2018,Husanu2020,Yukawa2016,Lee2014,Lin2014,Cancellieri2016}
We refer to this extended model as the ``doped \F\ solid''.
For this model, we derive the Fan-Migdal self-energy, and use it to obtain the electron spectral
function within both Dyson's equation and the second-order cumulant expansion
method. For each approach, we analyze the quasiparticle (QP) band structure, the phonon satellites, 
and the mass enhancement, and we identify advantages and shortcomings.
In particular, we show that both approaches yield anomalous electron mass enhancements
at finite Fermi levels. This enhancement is so strong that the band curvature is inverted
in a large region of the phase diagram, leading to a breakdown of the QP picture.
This failure is more pronounced in the cumulant approach.

This manuscript is organized as follows: In Sec.~\ref{sec.theory}, we 
formally introduce the doped \F\ solid, the Fan-Migdal self-energy, and how 
to obtain the spectral function within either Dyson's method or the 
second-order cumulant expansion. In Sec.~\ref{sec.res.undoped}, we review 
the classic \F\ polaron problem as the empty-band limit of the doped \F\ 
solid. We show that the empty-band model fails to reproduce the correct 
energetic ordering of bands and satellites that is observed in experiments.
This shortcoming is remedied in Sec.~\ref{sec.res.doped}, where we 
introduce free carriers and a finite Fermi level in the model. In this case
we only consider band filling effects, without taking into account the 
screening of the \F\ interaction by the free carriers. This scenario is 
relevant for experiments in the anti-adabatic regime 
($\hbar\omega_{\rm ph}/E_{\rm F}\gg 1$). In Sec.~\ref{sec.res.screened}, we
include both the effect of band filling and free-carrier screening, and 
derive semi-analytical self-energy and spectral functions. This more 
accurate model is found to capture most of the features observed in ARPES 
experiments in doped oxides. For easier orientation within this manuscript, 
we reference all equations for the QP energy and effective mass for the
three considered scenarios in table~\ref{tab1}.
\nk{Section~\ref{sec.qp.weight} connects our Dyson and cumulant spectra
to experimental data by comparing calculated and measured quasi-particle
weights.} In Sec.~\ref{sec.conc} we 
summarize our results and discuss the implications of our findings for 
\textit{ab initio} calculations of \F\ couplings. Lastly, we report details
of the derivations in the appendices.

\section{Model setup and general expressions for the self-energy and the spectral function}\label{sec.theory}

\subsection{The doped Fr{\"o}hlich solid}

The \F\ Hamiltonian for electrons coupled to dispersionless LO phonons is given by:\cite{Giustino2017}
 \begin{multline}
 \hat{H}=\sum_\textbf{k}\epsilon_\textbf{k}\,
 \hat{c}^\dagger_\textbf{k}\hat{c}_\textbf{k}
 +\hbar\omega_0\sum_\textbf{q}
 \Big(\hat{a}^\dagger_\textbf{q}\hat{a}_\textbf{q}+\frac12\Big)
 \\
 +N_p^{-1/2}\sum_{\textbf{k},\textbf{q}}g(q)\,
 \hat{c}^\dagger_{\textbf{k}+\textbf{q}}\hat{c}_\textbf{k}\,
 \big(\hat{a}^\dagger_\textbf{q}+\hat{a}_{-\textbf{q}}\big),
\end{multline}
where $\bf k$, $\bf q$, $\hat{c}_\textbf{k}$, and $\hat{a}_\textbf{q}$ are 
electron wavevectors, phonon wavevectors, fermion annihilation operators, 
and boson annihilation operators, respectively. The single-particle 
energies of the electrons are indicated by $\epsilon_\textbf{k}$, 
$\hbar\omega_0$ is the LO phonon energy, $g(q)$ with $q=|{\bf q}|$ is the 
\F\ matrix element, and $N_p$ is the number of unit cells in the 
Born-von-K{\'a}rm{\'a}n supercell.

In this model, the electron-electron interation is assumed to be already
taken into account by the effective mass $m_0$, and the electron band 
structure is simply given by $\epsilon_\textbf{k} = \hbar^2 k^2/2m_0$, with 
$k=|{\bf k}|$. Throughout this manuscript we consider the system at zero 
temperature, so that the electron occupations are described by the 
Heaviside function $f_\textbf{k}=\theta(k_{\rm F}-k)$, where $k_{\rm F}$ is
the Fermi wavevector. 

The matrix element of the \F\ interaction is given 
by:\cite{Verdi2015,Sjakste2015}
  \begin{equation}\label{eq.def.g2}
  g(q)=\frac{i}{q}
  \Bigg[\frac{4\pi\,\alpha\,\hbar(\hbar\omega_0)^{3/2}}
  {\Omega\,\sqrt{2m_0}} \Bigg]^{1/2},
  \end{equation}
where $\Omega=N_p\,\Omega_\text{UC}$ is the volume of the crystal cell
consisting of $N_p$ unit cells with volume $\Omega_\text{UC}$, and the 
strength of the interaction is quantified by the dimensionless \F\,
coupling constant:
\begin{equation}
	\alpha=\frac{e^2}{4\pi\epsilon_0\,\hbar}
	\sqrt{\frac{m_0}{2\hbar\omega_0}}
	\left(\frac{1}{\varepsilon_\infty}-\frac{1}{\varepsilon_0}\right).
\end{equation}
In this expression, $\epsilon_0$ is the vacuum permittivity, and 
$\varepsilon_0$ and $\varepsilon_\infty$ are the static and high-frequency 
dielectric constant of the undoped crystal.  The matrix element provided by 
Eq.~\eqref{eq.def.g2} describes the probability amplitude for an electron 
in the initial electronic state with wavevector ${\bf k}$ to be scattered 
into the final state with wavevector ${\bf k}+{\bf q}$ by an LO phonon of 
wavevector ${\bf q}$. The characteristic singularity at $q=0$ corresponds 
to the onset of a macroscopic polarization in the crystal, accompanied by a 
uniform electric field. In the polaron literature it is common to 
distinguish weak-coupling, intermediate coupling, and strong coupling 
depending on the value of $\alpha$.\cite{Devreese2009} Although this 
separation is somewhat arbitrary, the onset of strong coupling is usually 
placed at $\alpha = 6$ for reasons that will become clear in 
Sec.~\ref{sec.res.undoped}.

In the presence of free carriers, the \F\ interaction described by Eq.~\eqref{eq.def.g2}
is weakened by the electronic screening.\cite{Verdi2017} To be consistent with the
parabolic electron bands employed in the \F\ model, we describe this screening using the
Lindhard dielectric function $\varepsilon(q,\omega)$, i.e. the dielectric function
of the electron gas in the random-phase approximation.\cite{Hedin1969}

The dielectric function in the random-phase approximation is given by:
\cite{Hedin1969}
\begin{eqnarray}
   \varepsilon(q,\omega)
   &=&1+r_s\left(\frac{4}{9\pi}\right)^{1/3}\frac{1}{\pi}\frac{1}{(q/k_{\rm F})^3} \times \nonumber \\
   &\times& \left[ 2q/k_{\rm F}+f\left(q/k_{\rm F}+\frac{(\omega+i\eta)/E_{\rm F}}{q/k_{\rm F}}\right)
      \right.
     \nonumber \\
   &_& \left. +f\left(q/k_{\rm F}-\frac{(\omega+i\eta)/E_{\rm F}}{q/k_{\rm F}}\right) \right],  
    \label{eq.lindhard}
\end{eqnarray}
where $E_{\rm F}$ is the Fermi energy measured from the band bottom, $\eta$ is a positive
infinitesimal, and the function $f$ is given by $f(z)=(1-z^2/4) \log[(z+2)/(z-2)]$. The quantity
$r_s$ in Eq.~\eqref{eq.lindhard} is the Wigner-Seitz radius of the electron gas, i.e. the
radius of a sphere that contains one electron on average. It is given by:\cite{Mahan1981}
\begin{equation}
   r_s=\frac{m_0}{a_0\,\varepsilon_\infty}\left(\frac{3}{4\pi\,n}\right)^{1/3},
\end{equation}
where $n$ is the density of free carriers, and $a_0$ the Bohr radius\nk{, and 
$\varepsilon_\infty$ is again} the high-frequency dielectric constant of the semiconductor in the
absence of free carriers. This scaling is needed so that the Lindhard function describes free
carriers within the dielectric environment of the semiconductor, as opposed to the standard
electron gas in a metal.  The screened \F\ matrix element is then obtained via:\cite{Mahan1981,Verdi2017}
\begin{equation}\label{eq.def.gscreened}
  g^{\rm scr}(q) = \frac{g(q)}
  {\varepsilon(q,\omega_0)}.
\end{equation}
This equation states that the bare electron-phonon interaction is screened by \textit{both} the
dielectric constant of the semiconductor without free carriers (this effect is 
included in $g$), and the metallic screening provided by the free carriers, embedded in the
dielectric continuum of the semiconductors (this effect is included in $\varepsilon$).
A detailed derivation of Eq.~\eqref{eq.def.gscreened} can be found in Sec.~6.3 of Ref.~\citenum{Mahan1981}.

We note that, in Eq.~\eqref{eq.def.gscreened}, we evaluate the Lindhard function at 
the phonon frequency $\omega_0$. This is a reasonable approximation that is necessary to
keep the problem tractable. A complete calculation including the frequency dependence of
$\varepsilon(q,\omega)$ would introduce additional poles in the self-energy, and would
require us to take into account phonon-plasmon polaritons. We have not explored these avenues given
the complexity of the formalism.

\subsection{Dyson's equation approach}

We describe the many-body band structure of the doped \F\ solid by calculating the electron 
spectral function:
  \begin{equation}\label{eq.def.spec}
  A_k(\omega)
  =\frac{1}{\pi}\left|\text{Im}\, G_k(\omega)\right|.
  \end{equation}
This function represents the momentum-resoved density of states and it is accessible via
ARPES experiments.\cite{Damascelli2003,Abrikosov1975} To obtain $A_k(\omega)$, we 
evaluate the interacting electron Green's function $G_k(\omega)$ of the system. Both the
spectral function and the Green's function depend only on the absolute value of the electron
wavevector as the system is isotropic.

In the Dyson equation approach, the Green's function is evaluated as $G = G_0 + G_0\Sigma G$,
where $G_0$ is the non-interacting Green's function, and $\Sigma$ is the self-energy.
This equation leads to the standard expression:
 \begin{equation}\label{eq.dyson}
 G_k(\omega)=
 \left[ \hbar\omega-\epsilon_k-\Sigma_k(\omega) \right]^{-1}.
 \end{equation}
Here, the Green's function and the self-enery are both retarded. The same results
would be obtained using the time-ordered version of both quantities.
By combining Eqs.~\eqref{eq.def.spec} and \eqref{eq.dyson} the spectral function can be
expressed directly in terms of the self-energy:
  \begin{equation}\label{eq.dm.spec}
  A_k(\omega)=\frac{\nk{-}1}{\pi}
  \frac{\text{Im}\,\Sigma_k(\omega)}
  {\left[
    \hbar\omega-\epsilon_k-\text{Re}\,\Sigma_k(\omega)
  \right]^2
  +\left[\text{Im}\,\Sigma_k(\omega)\right]^2}.
  \end{equation}
The electron addition/removal energies correspond to the poles of the
Green's function, and are usually determined by setting to zero the 
denominator of Eq.~\eqref{eq.dm.spec} under the assumption that the 
imaginary part of the self-energy \nk{and its frequency dependence can be 
neglected near the poles}. By calling these poles $E_k$, we have:
  \begin{equation}\label{eq.dm.ek}
  E_k=\epsilon_k + \text{Re}\,\Sigma_k(E_k),
  \end{equation}
or, linearized around $\epsilon_k$:
  \begin{equation}\label{eq.dm.ek2}
  E_k=\epsilon_k +Z_k \,\text{Re}\,\Sigma_k(\epsilon_k),
  \end{equation}
where the QP renormalization factor $Z_k$ is given by:
\begin{equation}\label{eq.dm.zk}
  Z_k= \left[
  1-\frac{1}{\hbar}
  \frac{\partial\,\text{Re}\,\Sigma(\omega)}
  {\partial\omega}
  \right]^{-1}_{\omega=E_k/\hbar}.
  \end{equation}
This quantity represents the spectral weight of the QP peak,
and $1-Z_k$ is the spectral weight transferred to the incoherent satellite structure,
i.e. the phonon sidebands schematically illustrated in Fig.~\ref{fig1}.

The evaluation of the self-energy requires the summation over all possible
connected electron-phonon Feynman diagrams. This summation can be performed
numerically using the diagrammatic Monte Carlo method,\cite{Prokofev1998} 
as it has been demonstrated for the original (undoped) \F\ 
model.\cite{Mishchenko2000} Here, we are interested in developing analytic
and semi-analytic solutions, therefore we truncate the expansion to the 
first-order diagram, consisting of a single electron line and a single 
phonon line connected by the electron-phonon matrix elements at the two 
ends. This choice leads to the Fan-Midgal 
self-energy:\cite{Migdal1958,Giustino2017}
  \begin{multline}\label{eq.sigma.t}
  \Sigma_k(\omega)=\frac{N_p}{\hbar}
  \int_\text{BZ}\frac{d\textbf{q}}{\Omega_\text{BZ}}
  \Bigg[
  \frac{|g(q)|^2\,f_{\textbf{k}+\textbf{q}}}
  {\omega-\epsilon_{\textbf{k}+\textbf{q}}/\hbar
  +\omega_0\nk{+}i\eta}\\+\frac{|g(q)|^2\,(1-f_{\textbf{k}+\textbf{q}})}
  {\omega-\epsilon_{\textbf{k}+\textbf{q}}/\hbar-\omega_0\nk{+}i\eta}
  \Bigg].
  \end{multline}
This self-energy describes the electron-phonon interaction to second order
in the atomic displacement, as can be seen from the fact that the \F\
matrix element appears as $g^2$. To the same order in perturbation 
theory, there exists an additional contribution to the self-energy, the 
Debye-Waller term.\cite{Allen1976,Allen1978} The Debye-Waller self-energy 
plays an important role in the calculation of phonon-induced band gap 
renormalization in semiconductors and insulators.\cite{Marini2008,
Giustino2010,Antonius2014} In the case of the \F\ model considered here, 
the Debye-Waller self-energy vanishes identically, as we show in 
Appendix~\ref{sec.debye}.

\subsection{Cumulant expansion approach}

A promising strategy to include higher-order electron-phonon diagrams 
beyond the Fan-Migdal self-energy is provided by the cumulant expansion 
formalism.\cite{Langreth1970,Hedin1980,Aryasetiawan1996,Lischner2013,
Kas2014, Caruso2015a,Caruso2015b,Caruso2016a,Caruso2016b,Gumhalter2016,
Zhou2018b,Tzavala2020, Story2014,Verdi2017,Caruso2018,Riley2018,Nery2018,
Zhou2019,Antonius2020,Zhou2020}
\nk{Owing to its roots in the description of deep-lying core states, the 
cumulant is \textsl{a priori} defined in terms of the lesser and greater
self-energy, clearly separating electron and hole states. Later adaptations
to states near the Fermi level include the introduction of the retarded
cumulant.\cite{Kas2014} In this manuscript, we follow the original 
definition of the cumulant expansion, treating electrons and holes 
separately.\cite{Gumhalter2016}

The interacting Green's function in the time domain is obtained as} the 
product of the non-interacting Green's function and the time-evolution 
operator:\cite{Langreth1970}
  \begin{equation}\label{eq.cum1}
  G^\gtrless_k(t,t')=
  G^\gtrless_{0,k}(t,t')\,e^{C^\gtrless_k(t-t')},
  \end{equation}
where $t,t'$ are time variables, and hole or electron QPs are described 
separately via the lesser ($<$) or greater ($>$) Green's function. The 
exponential represents the time-evolution operator, and $C^\gtrless_k$ is 
the cumulant function.

We expand the exponential in Eq.~\eqref{eq.cum1} and compare the 
term linear in $C_k^\gtrless(t-t')$ to the expansion of the 
Dyson equation:\cite{Aryasetiawan1996}
  \begin{multline}\label{eq.cum2}
  G^\gtrless_{0,k}(t,t')C^\gtrless_k(t-t')\\ 
  =\frac{1}{2\pi}\int_{-\infty}^\infty d\omega\, G^\gtrless_{0,k}(\omega)
  \Sigma^\gtrless_k(\omega) G^\gtrless_{0,k}(\omega)\,e^{-i\omega t}.
  \end{multline}
Using the definition of the lesser (greater) non-interacting Green's 
function, 
  \begin{equation}
  G_k^\gtrless(t,t')
  =\mp\frac{i}{\hbar}\theta(\pm t\mp t')
  e^{-\frac{i}{\hbar}(\epsilon_k\mp i\eta)(t-t')},
  \end{equation}
inside Eq.~\eqref{eq.cum2}, the cumulant function can be expressed in 
terms of the same self-energy employed in Dyson's equation approach:
  \begin{eqnarray}\label{eq.cum3}
  C^\gtrless_k(t,t')=\frac{\mp1}{\pi\,\hbar} \int d\omega\,
   \text{Im}\Sigma^\gtrless_k(\epsilon_k\pm\omega)
   \frac{1\mp i\omega t -e^{\mp i\omega t}}{\omega^2}.\nonumber\\
  \end{eqnarray}
If we use the Fan-Migdal self-energy, the cumulant function will also contain electron-phonon
interactions to second order in the atomic displacements. The advantage of the cumulant method
is that, when the approximate cumulant function given by Eq.~\eqref{eq.cum3} is used inside
Eq.~\eqref{eq.cum1}, the exponentiation or ``cumulant resummation''\cite{Mahan1981} 
generates an infinite series of terms. This series contains both non-crossing and crossing 
electron-phonon Feynman diagrams.\cite{Hedin1980}

The lesser and greater self-energies appearing in Eq.~\eqref{eq.cum3} are
given by:
  \begin{eqnarray}\label{eq.sigma.l}
  &&\Sigma^<_k(\omega)=\frac{N_p}{\hbar}
  \int_\text{BZ}\frac{d\textbf{q}}{\Omega_\text{BZ}}
  \frac{|g(q)|^2\,f_{\textbf{k}+\textbf{q}}}
  {\omega-\epsilon_{\textbf{k}+\textbf{q}}/\hbar +\omega_0-i\eta}, \\
  \label{eq.sigma.g} && \Sigma^>_k(\omega)=\frac{N_p}{\hbar}
  \int_\text{BZ}\frac{d\textbf{q}}{\Omega_\text{BZ}} 
  \frac{|g(q)|^2\,(1-f_{\textbf{k}+\textbf{q}})}
  {\omega-\epsilon_{\textbf{k}+\textbf{q}}/\hbar -\omega_0+i\eta}.
  \end{eqnarray}
These self-energies are simply related to the retarded self-energy of Eq.~\eqref{eq.sigma.t}
by $\Sigma_k = \nk{\left(\Sigma^<_k\right)^*+\Sigma^>_k}$.

In order to gain insight into the structure of the spectral function obtained from
the cumulant expansion, it is convenient to express Eqs.~\eqref{eq.cum1} and \eqref{eq.cum3}
in the frequency domain. The result is:
  \begin{eqnarray}
  A^{\gtrless}_k &=&A^{\gtrless}_{{\rm QP},k}
  +A_{\text{QP},k}^\gtrless*A_{\text{S},k}^\gtrless \nonumber \\
  &+&
  \frac{1}{2}A^\gtrless_{\text{QP},k} *A^\gtrless_{\text{S},k}*A_{\text{S},k}^\gtrless
  +\cdots, \label{eq.conv}
  \end{eqnarray}
where $*$ denotes a convolution in frequency space. A detailed derivation 
of this result can be found in Ref.~\citenum{Aryasetiawan2000}. The 
functions $A_{\text{QP},k}^\gtrless(\omega)$ and 
$A_{\text{S},k}^\gtrless(\omega)$ are given by:
  \begin{eqnarray}\label{eq.qp}
  &&A_{\text{QP},k}^\gtrless(\omega)
  =\frac{Z_k^\gtrless}{\pi}
  \begin{pmatrix} 1-f_k\\ f_k \end{pmatrix}\\ 
  &&\times\frac{\text{Im}\,\Sigma^\gtrless_k(\epsilon_k)
  \cos\alpha^\gtrless_k
  -\left[\omega-\epsilon_k
  -\text{Re}\,\Sigma^\gtrless_k(\epsilon_k)\right]
  \sin\alpha^\gtrless_k}
  {\left[\omega-\epsilon_k
  -\text{Re}\,\Sigma^\gtrless_k(\epsilon_k)\right]^2
  +\left[\text{Im}\,\Sigma^\gtrless_k(\epsilon_k)
  \right]^2},\qquad\nonumber
  \end{eqnarray}
  \begin{equation}\label{eq.s}
  A_{\text{S},k}^\gtrless(\omega)=
  \frac{\mp\text{Im}\,\Sigma^\gtrless_k(\epsilon_k+\omega)
  -\left(\mp\text{Im}\,\Sigma^\gtrless_k(\epsilon_k)
  \mp\omega\,\alpha_k^\gtrless\right)}
  {\pi\,\hbar^2\omega^2},
  \end{equation}
where the quantities $\alpha_k^\gtrless$ and $Z_k^\gtrless$ are defined as:
  \begin{equation}
  \alpha^\gtrless_k= \left.\frac{\partial\,
  \text{Im}\,\Sigma^\gtrless_k(\omega)}
  {\partial\omega}\right|_{\omega=\epsilon_k/\hbar},
  \end{equation}
  \begin{equation}\label{eq.ce.zk}
  Z^\gtrless_k=\exp
  \left( \frac{\partial\, \text{Re}\,\Sigma^\gtrless_k(\omega)}
  {\partial\omega}\right)_{\omega=\epsilon_k/\hbar}.
  \end{equation}
The first term on the r.h.s. of Eq.~\eqref{eq.conv} represents the QP 
peak, and corresponds to a Fano lineshape.
The QP peak is found at the energy:
  \begin{equation}\label{eq.ce.ek}
  E^{\gtrless}_k= \epsilon_k+
  \text{Re}\,\Sigma^\gtrless_k(\epsilon_k).
  \end{equation}
Successive terms of the series expansion in Eq.~\eqref{eq.conv} represent a sequence of satellites, 
one per convolution. Higher-order convolutions correspond to weaker satellites located farther away 
from the QP peak. In practice, the first two to three satellites
carry the majority of the spectral weight of the incoherent part and are the features
usually resolved in experiments.

\nk{
One question that often arises in the cumulant expansion literature is whether
one should use the lesser and greater self-energy, whereby electrons and holes are
described separately,\cite{Gumhalter2016,Aryasetiawan2000} or else one should 
use the retarded self-energy, whereby electrons and holes are described at the same 
time.\cite{Kas2014}
	
If the cumulant is used to describe electron or hole states away from the 
Fermi level, the lesser (greater) Green's function only depends on the lesser
(greater) self-energy. It has been shown that this picture can be extended to 
states near the Fermi surface.\cite{Aryasetiawan1996,Lischner2013}
The retarded cumulant introduced in Ref.\citenum{Kas2014}
is designed to describe emission and absorption processes simultaneously. 
The main difference between the retarded cumulant and the lesser/greater 
self-energy approach used in the present work lies in the description 
of the satellites.

The satellite function depends exclusively on the imaginary part of 
the self-energy, which is a quantity that can easily be separated into 
contributions arising from absoption and emission. In particular, we have:
  \begin{equation}
  \text{Im}\left[\Sigma^<(\omega)\right]\ne 0 \quad\text{only if} \quad \omega<-\omega_0,
\end{equation}
which implies that the lesser self-energy can only give rise to hole 
satellites at $\omega<-\omega_0$, i.e. \textsl{below} the quasiparticle peak. 

Conversely, for the greater self-energy, we have
\begin{equation}
	\text{Im}\left[\Sigma^>(\omega)\right]\ne 0 \quad\text{only if} \quad \omega>\omega_0,
\end{equation}
causing electron satellites \textsl{above} the quasiparticle peak. 
Crucially, the shape and magnitude of the lesser and greater 
satellite functions are completely independent.
	
Given this premise, the difference between different 
cumulant approaches can be understood as follows:
The retarded cumulant employs both the lesser and greater self-energy 
at all $k$ points, causing satellites to appear above and below the 
quasiparticle peaks throughout the band structure. Conversely, in the lesser/greater
self-energy approach, hole satellites are confined to states $k<k_{\rm F}$, 
and electron satellites to states $k>k_{\rm F}$.
	
To the best of the authors' knowledge, in ARPES 
measurements\cite{Santander-Syro2011,WangZ2016,Riley2018,Meevasana2011} 
phonon satellites are only observed for wavevectors smaller than the Fermi wavevector. 
Since in the retarded cumulant, the satellites are found to disperse beyond the Fermi wavevector, 
the present approach appears more suitable to model existing ARPES data.}

\nk{To conclude this section, we briefly note the main differences between the Dyson's
approach and the second-order cumulant:} 
(i) Dyson's approach using the Fan-Migdal 
self-energy leads to one QP peak and one satellite, while the cumulant 
approach leads to one QP peak and a series of satellites of decreasing 
intensity; (ii) in Dyson's approach, the self-energy is evaluated at the 
QP energy $E_k$, while in the cumulant approach the self-energy is 
evaluated at the non-interacting energy $\epsilon_k$ 
(``on the mass shell''). This latter difference leads to different QP 
energies: Dyson's method contains the renormalization factor $Z_k$ 
[see Eq.~\eqref{eq.dm.ek2}], but the cumulant approach does not 
[see Eq.~\eqref{eq.ce.ek}]. This inconsistency is reflected in the QP 
effective masses, as we discuss in Sec.~\ref{sec.res.undoped}

\section{Results}\label{sec.results}

\subsection{Single electron in the conduction band}\label{sec.res.undoped}

We start by considering the case of a single electron added to an otherwise
empty conduction band, which corresponds to the well-known \F\ polaron 
problem.\cite{Frohlich1950} The self-energy for this case is obtained by 
setting $f_k = 0$ for all electron wavevectors $k$. As a result, the lesser
self-energy in Eq.~\eqref{eq.sigma.l} vanishes identically, and the 
retarded self-energy in Eq.~\eqref{eq.sigma.t} \nk{is equal to the greater
self-energy in Eq.~\eqref{eq.sigma.g}}. After performing a change of 
integration variables to spherical coordinates, Eq.~\eqref{eq.sigma.g} can 
be integrated analytically to yield:
  \begin{equation}\label{eq.sigma.g.undoped}
  \Sigma^>_k(\omega)
  =-i\frac{\alpha\,(\hbar\omega_0)^{3/2}}{2\pi\,\sqrt{\epsilon_k}}
  \log\frac{\sqrt{\hbar\omega-\Omega_0}+\sqrt{\epsilon_k}}
  {\sqrt{\hbar\omega-\Omega_0}-\sqrt{\epsilon_k}},
  \end{equation}
where $\Omega_0=\hbar(\omega_0-i\eta)$. This result was also 
derived, among others, in Refs.~\citenum{Schlipf2018, Mahan1981}. Some key 
steps of the derivation are reported in Appendix~\ref{app.steps.undoped}.

The real and imaginary parts of the self-energy are shown in 
Figs.~\ref{fig2}(a) and (b), respectively. As the real part of the 
self-energy is negative everywhere [see Fig.~\ref{fig2}(a)]
the QP energy near the bottom of the conduction band lies below the 
non-interacting energy. The physical interpretation of this result is 
that the phonon cloud tends to stabilize the electron, precisely as it 
happens when a polaron is formed.\cite{Franchini2021} This qualitative 
trend holds for both the Dyson's approach and the cumulant approach, 
as it can be seen in the spectral functions reported 
in Figs.~\ref{fig2}(c), (d), and (e), (f), respectively. 

The imaginary part of the self-energy vanishes identically for frequencies 
$\omega$ below the threshold $\epsilon_k+\hbar\omega_0$, as it can be seen 
in Fig.~\ref{fig2}(b). The interpretation of this behavior is that the 
electron does not have sufficient energy to emit a phonon, therefore its 
lifetime is infinite and ${\rm Im}\,\Sigma = 0$. This effect is also seen 
in the spectral functions, Figs.~\ref{fig2}(c) - (f), which exhibit
sharp QP peaks for energies within $\hbar\omega_0$ from the band bottom.

Despite sharing the same self-energy, the Dyson and cumulant approaches 
differ considerably in the QP energies and effective masses. In the Dyson 
approach, the QP energy is defined by:
  \begin{equation}\label{eq.ek.undoped}
  E_k=\epsilon_k
  +\frac{\alpha\,(\hbar\omega_0)^{3/2}}{2\pi\,\sqrt{\epsilon_k}}
  \arg\frac{\sqrt{E_k-\Omega_0^*}+\sqrt{\epsilon_k}}
  {\sqrt{E_k-\Omega_0^*}-\sqrt{\epsilon_k}}.
  \end{equation}
This expression does not lead to a general analytic expression for $E_k$, 
but the QP energy and mass at the band bottom ($k=0$) have simple 
expressions, see e.g. Section 7.1.1 of Ref.~\citenum{Mahan1981}:
  \begin{equation}\label{eq.en.dyson.undop}
  \frac{E_0}{\hbar\omega_0}=-\frac{\alpha}
  {\sqrt{1-E_0/\hbar\omega_0}},
  \end{equation}
and
  \begin{equation}\label{eq.mass.dyson.undop}
  \frac{m^{*}}{m_0}=
  \frac{\displaystyle 1+\alpha/2}{\displaystyle 1+\alpha/3}.
  \end{equation}
The weak coupling ($\alpha\ll 1$) expansion of Eq.~\eqref{eq.en.dyson.undop}
can be obtained by writing the solution $E_0$ as a continued fraction and 
then taking the limit of small $\alpha$:
  \begin{equation}\label{eq.en.dyson.undop.expand}
  \frac{E_0}{\hbar\omega_0} = -\alpha+\frac{\alpha^2}{2}
  -\frac{5}{8}\alpha^3+\mathcal{O}(\alpha^4).
  \end{equation}
Similarly, the weak-coupling expansion of Eq.~\eqref{eq.mass.dyson.undop} 
is
  \begin{equation}\label{eq.mass.dyson.undop.expand}
  \frac{m^{*}}{m_0}=
   1+\frac{\alpha}{6}-\frac{\alpha^2}{18}+\mathcal{O}(\alpha^3).
  \end{equation}
The effective mass renormalization in the Dyson approach is the same as 
that obtained within  Brillouin-Wigner perturbation theory applied to the 
\F\ polaron problem.\cite{Mahan1981}

In the case of the cumulant approach, the QP energy is given by:
  \begin{equation}\label{eq.ek.cumul.undop}
  E^>_k=\epsilon_k
  +\frac{\alpha\,(\hbar\omega_0)^{3/2}}{2\pi\,\sqrt{\epsilon_k}}
  \arg\frac{\sqrt{\epsilon_k-\Omega_0}+\sqrt{\epsilon_k}}
  {\sqrt{\epsilon_k-\Omega_0}-\sqrt{\epsilon_k}}.
  \end{equation}
By taking the limit of small $k$, we obtain the standard result for the QP 
energy at the band bottom,
  \begin{equation}\label{eq.en.cumul.undop}
  \frac{E_0^>}{\hbar\omega_0}=-\alpha,
  \end{equation}
which is valid at all $\alpha$. The corresponding effective mass is: 
  \begin{equation}\label{eq.mass.cumul.undop}
  \frac{m^{*,>}}{m_0}=\frac{1}{1-\alpha/6}
  = 1 + \frac{\alpha}{6}+\frac{\alpha^2}{36}
  +\mathcal{O}(\alpha^3).
  \end{equation}
These last two results coincide with what one obtains by performing
Rayleigh-Schr{\"o}dinger perturbation theory on the \F\ polaron 
problem.\cite{Schiff1955}

It is instructive to compare Eqs.~\eqref{eq.en.dyson.undop}, 
\eqref{eq.mass.dyson.undop.expand}, \eqref{eq.en.cumul.undop}, and 
\eqref{eq.mass.cumul.undop} with calculations based on Feynman's path 
integral approach to the \F\ polaron problem.\cite{Feynman1954} Feynman's 
approach is considered to be the most accurate in describing the undoped \F\ 
model, and agrees closely with advanced diagrammatic Monte Carlo 
calculations.\cite{Mishchenko2000} In this approach, the QP energy and 
mass at weak coupling are found to be:
  \begin{equation}
  \frac{E_0}{\hbar\omega_0}=-\alpha
  -\frac{1}{81}\alpha^2+\mathcal{O}(\alpha^3).
  \end{equation}
  \begin{equation}
  \frac{m^*}{m_0}=1+\frac{\alpha}6+0.025\alpha^2+\mathcal{O}(\alpha^3).
  \end{equation}
These two results show that, at \textit{weak} coupling ($\alpha \ll 1$), 
\textit{both} the Dyson approach and the cumulant approach yield QP energies
and effective masses that agree with Feynman's path integral calculation to 
first order in the coupling strength $\alpha$. A detailed comparison between
these three approaches to the undoped \F\ model is shown in Fig.~\ref{fig3}.
This comparison shows that, while the three approaches agree at weak 
coupling, there exist significant differences for larger values of $\alpha$.
In particular, the cumulant method yields a QP energy that remains closer 
to the Feynman result up to intermediate coupling strengths ($\alpha=6$), 
while the Dyson approach deviates from Feynman's already at moderate 
coupling. On the other hand, the cumulant approach yields an unphysical 
divergence of the effective mass at intermediate coupling (singularity at 
$\alpha=6$ and change of sign beyond this point), while the mass in Dyson's
method remains finite.

Based on the comparison between QP energies, recently it has been argued
that the cumulant method provides a better description of polarons than 
Dyson's approach.\cite{Nery2018}
However, Fig.~\ref{fig3} clearly shows that the effective mass (and 
by extension the band structure) in the cumulant approach is not reliable 
at intermediate coupling. This point is further corroborated by a close 
inspection of the spectral functions in Fig.~\ref{fig2}(e): As a result of 
a logarithmic singularity in the self-energy [Eq.~\eqref{eq.sigma.g.undoped}],
the QP energy diverges when $\epsilon_k=\hbar\omega_0$, and the spectral 
function exhibits unphysical vertical streaks. 

Moving to the polaron satellites, we see from Figs.~\ref{fig2}(c) and (e) 
that both the Dyson's approach and the cumulant approach exhibit 
satellites states. As already discussed in numerous 
reports,\cite{Aryasetiawan1996,Lischner2013,Kas2014,Lischner2014,
Story2014,Caruso2015a,Caruso2015b,Lischner2015,Zhou2015,Caruso2016b,
Gumhalter2016,Vigil-Fowler2016,Verdi2017,Nery2018,Zhou2018b,Zhou2019,
Antonius2020,Tzavala2020,Zhou2020,Chang2021,Zhou2021} Dyson's approach yields only one 
satellite, to leading order located at a binding energy of 
$(1+\alpha)\hbar\omega_0$ from the QP peak, see Fig.~\ref{fig2}(c).
On the other hand, the cumulant method correctly yields multiple 
satellites which are separated from the QP peak by integer multiples of 
the boson energy $\hbar\omega_0$. Thus, the cumulant method is superior in 
the description of satellite features, as anticipated.

One last issue that deserves attention is the location of the satellites 
with respect to the QP band. Both Dyson's method and the cumulant approach
yield satellites located at \textit{higher} energy than the QP band [see
Figs.~\ref{fig2}(c) and (e)] \nk{when applied to the empty-band system}. 
However, in ARPES experiments satellites are observed \textit{below} the QP
band, as schematically illustrated in Fig.~\ref{fig1}. This discrepancy 
has to do with the fact that ARPES probes \textit{occupied} electronic 
states, while the empty-band \F\ model describes \textit{unoccupied} 
states. It is clear that a correct description of polaron physics as 
probed in ARPES experiments necessitates the study of a \textit{doped} \F\ 
solid. The following sections are devoted to the doped model.

\subsection{Finite Fermi level in the conduction band}\label{sec.res.doped}

Now we consider the case of partially occupied conduction band with a 
Fermi energy $E_{\rm F}>0$. The self-energy for this case is obtained by 
setting $f_\textbf{k}=\theta(k_{\rm F}-k)$ in Eqs.~\eqref{eq.sigma.l} and 
\eqref{eq.sigma.g}. In this section, we ignore free-carrier screening, which
will be included in Sec.~\ref{sec.res.screened}. This approximation is 
meaningful to describe the anti-adiabatic regime, where the Fermi level is 
much smaller than the characteristic phonon energy, 
$E_{\rm F}\ll\hbar\omega_0$.

After carrying out the integrals in Eqs.~\eqref{eq.sigma.l} and 
\eqref{eq.sigma.g} explicitly, we obtain the following self-energies. For 
completeness, key steps of the derivation are provided in 
Appendix~\ref{app.steps.undoped}. The lesser self-energy, which describes 
electron removal processes, is given by
  \begin{eqnarray}\label{eq.dop.sigma.l}
  &&\Sigma_k^<(\omega) =-\frac{\alpha\,(\hbar\omega_0)^{3/2}}
  {2\pi\,\sqrt{\epsilon_k}}
  \Bigg[ L\left(\sqrt{E_{\rm F}/\epsilon_k}, 
  \sqrt{(\hbar\omega+\Omega_0)/\epsilon_k}\right) 
  \nonumber \\ 
  &&+\log\frac{\hbar\omega+\Omega_0-E_{\rm F}} 
  {\hbar\omega+\Omega_0-\epsilon_k}
  \log\left|\frac{\sqrt{E_{\rm F}}+\sqrt{\epsilon_k}}
  {\sqrt{E_{\rm F}}-\sqrt{\epsilon_k}}\right| \Bigg]
  -\text{Re}\,\Sigma^<_{k_{\rm F}}(E_{\rm F}),\nonumber\\
  \end{eqnarray}
with
  \begin{eqnarray}\label{eq.dop.sigma.l.ef}
  \Sigma^<_{k_{\rm F}}(E_{\rm F}) = 
  -\frac{\alpha\,(\hbar\omega_0)^{3/2}}{2\pi\,\sqrt{E_{\rm F}}}
  &\Bigg[&
    \text{Li}_2\frac{2\,\sqrt{E_{\rm F}}}
    {\sqrt{E_{\rm F}}+\sqrt{E_{\rm F}+\Omega_0}}\nonumber\\
    &+&\text{Li}_2\frac{2\,\sqrt{E_{\rm F}}}
    {\sqrt{E_{\rm F}}-\sqrt{E_{\rm F}+\Omega_0}}
  \Bigg].\hspace{8mm}
  \end{eqnarray}
In these expressions, the auxiliary function $L$ is defined as:
  \begin{eqnarray}
  L(z_1,z_2)&=&\text{Li}_2\frac{1+z_1}{1+z_2}
  +\text{Li}_2\frac{1-z_1}{1+z_2} \nonumber \\
  &-&\text{Li}_2\frac{1+z_1}{1-z_2} -\text{Li}_2\frac{1-z_1}{1-z_2},
  \hspace{1cm}
  \end{eqnarray}
Li$_2$ denotes the dilogarithm function, and $z_1,z_2$ are
complex-valued parameters.
The greater self-energy, which describes electron addition processes, is 
found to be:
  \begin{eqnarray}\label{eq.dop.sigma.g}
  \Sigma_{k}^>(\omega) &=&\frac{\alpha\,(\hbar\omega_0)^{3/2}}
  {2\pi\,\sqrt{\epsilon_k}}
  \Bigg[  L\left(\sqrt{E_{\rm F}/\epsilon_k},
  \sqrt{(\hbar\omega-\Omega_0)/\epsilon_k}\right)
  \nonumber \\  &+&
  \log\frac{\hbar\omega-\Omega_0-E_{\rm F}} 
  {\hbar\omega-\Omega_0+\epsilon_k}
  \log\left|\frac{\sqrt{E_{\rm F}}+\sqrt{\epsilon_k}}
  {\sqrt{E_{\rm F}}-\sqrt{\epsilon_k}}\right|
  \nonumber \\  &-&
  i\pi\log\frac{\sqrt{\hbar\omega-\Omega_0}+\sqrt{\epsilon_k}} 
  {\sqrt{\hbar\omega-\Omega_0}-\sqrt{\epsilon_k}} \Bigg]
  -\text{Re}\,\Sigma^>_{k_{\rm F}}(E_{\rm F}),\hspace{5mm}
  \end{eqnarray}
with
  \begin{eqnarray}\label{eq.dop.sigma.g.ef}
  &&\Sigma^>_{k_{\rm F}}(E_{\rm F}) = 
  -\frac{\alpha\,(\hbar\omega_0)^{3/2}}{2\pi\,\sqrt{E_{\rm F}}}
  \Bigg[i\pi\,\log\frac{\sqrt{E_{\rm F}-\Omega_0}+\sqrt{E_{\rm F}}}
    {\sqrt{E_{\rm F}-\Omega_0}-\sqrt{E_{\rm F}}}\nonumber \\
    &&-\text{Li}_2\frac{2\,\sqrt{E_{\rm F}}}
    {\sqrt{E_{\rm F}}+\sqrt{E_{\rm F}-\Omega_0}}
    -\text{Li}_2\frac{2\,\sqrt{E_{\rm F}}}
    {\sqrt{E_{\rm F}}-\sqrt{E_{\rm F}-\Omega_0}}
  \Bigg].\hspace{5mm}
  \end{eqnarray}
For metallic systems, Luttinger's theorem states that the volume of the 
Fermi surface does not change when adiabatically turning on many-body interactions in 
a non-interacting system.\cite{Luttinger1960} In the present model, the 
volume of the Fermi surface is determined by the Fermi wavevector 
$k_{\rm F}$, therefore Luttinger's theorem implies that $k_{\rm F}$ and 
hence the Fermi energy $E_{\rm F}$ should not be affected by the 
self-energy $\Sigma$. In order to \textit{enforce} this condition, 
we subtracted the constants defined in Eqs.~\eqref{eq.dop.sigma.l.ef} and 
\eqref{eq.dop.sigma.g.ef} from the lesser and greater self-energies
in Eqs.~\eqref{eq.dop.sigma.l} and \eqref{eq.dop.sigma.g}, 
respectively, so that ${\rm Re}\,\Sigma_{k_{\rm F}}(E_{\rm F}/\hbar)=0$.
One can verify that this choice leaves the QP energy at the 
Fermi level identical to the non-interacting energy, for both the Dyson's 
and cumulant approaches. 

Using Eqs.~\eqref{eq.dop.sigma.l} and \eqref{eq.dop.sigma.g}, the 
retarded self-energy is obtained as 
\nk{$\Sigma=\left(\Sigma^<\right)^*+\Sigma^>$}.
We note that Eq.~\eqref{eq.dop.sigma.g} correctly reduces to the 
corresponding equation for the undoped model, 
Eq.~\eqref{eq.sigma.g.undoped}, upon taking the limit 
$E_{\rm F}\rightarrow 0$.
The real and imaginary parts of the doped self-energy are shown in 
Figs.~\ref{fig4}(a) and (b). The real self-energy is now positive in the
range of occupied states, passes through zero at 
$(k=k_{\rm F},\hbar\omega=E_{\rm F})$ to ensure particle number 
conservation, and becomes negative for unoccupied states. The resulting 
QP peak shown in Figs.~\ref{fig4}(c) and (e) thus exhibits a 
higher effective mass than the bare electron. In fact, we find 
that the mass renormalization in the presence of doping is even more
pronounced than in the empty-band model, as we discuss below.

Turning to the imaginary self-energy shown in Fig.~\ref{fig4}(b), we note 
that the main peak structure in $\text{Im}\Sigma^<_k$ for occupied states,
$k<k_{\rm F}$, is now found at lower energies than the independent particle, 
causing the satellites in Figs.~\ref{fig4}(c) and (d) to appear below the 
QP peak. For empty states, the situation is similar to the 
discussion of the empty-band model in Sec.~\eqref{sec.res.undoped}, i.e. we
find satellite features above the QP dispersions.

In both Dyson's and cumulant approaches we find that QP
energy and effective mass are strongly doping-dependent. 
Starting with the Dyson approach, the dressed electron energy is found 
to be:
\nk{
  \begin{eqnarray}\label{eq.ek.dyson.dop}
  E_k &=&\epsilon_k-
  \frac{\alpha\,(\hbar\omega_0)^{3/2}}{2\pi\,\sqrt{\epsilon_k}}
  \text{Re}\Bigg[
    L\left(\sqrt{E_{\rm F}/\epsilon_k}, 
    \sqrt{(E_k+\Omega_0^*)/\epsilon_k}\right) \nonumber \\ 
    &-&L\left(\sqrt{E_{\rm F}/\epsilon_k}, 
    \sqrt{(E_k-\Omega_0)/\epsilon_k}\right) \nonumber \\
    &+&\log\frac{E_k+\Omega_0^*-E_{\rm F}}{E_k+\Omega_0^*-\epsilon_k}
    \log\left|\frac{\sqrt{E_{\rm F}}+\sqrt{\epsilon_k}} 
    {\sqrt{E_{\rm F}}-\sqrt{\epsilon_k}}\right| \nonumber \\ 
    &-&\log\frac{E_k-\Omega_0-E_{\rm F}}{E_k-\Omega_0+\epsilon_k}
    \log\left|\frac{\sqrt{E_{\rm F}}+\sqrt{\epsilon_k}}
    {\sqrt{E_{\rm F}}-\sqrt{\epsilon_k}}\right| \nonumber \\
    &+&i\pi\log\frac{\sqrt{E_k-\Omega_0}+\sqrt{\epsilon_k}}
    {\sqrt{E_k-\Omega_0}-\sqrt{\epsilon_k}} 
  \Bigg]
  -\text{Re}\,\Sigma_{k_{\rm F}}(E_{\rm F}).
  \end{eqnarray}
}
At the bottom of the conduction band $(k=0)$, this result yields
the following expressions for the QP energy and the effective mass:
\nk{
  \begin{eqnarray}\label{eq.en.dyson.dop}
  \frac{E_0}{\hbar\omega_0}&=&\frac{\alpha}{\pi}
  \text{Re}\Bigg[
    \frac{\sqrt{\hbar\omega_0}}{\sqrt{E_0+\Omega_0^*}}
    \log\frac{\sqrt{E_0+\W0^*}+\sqrt{\ef}}
    {\sqrt{E_0+\W0^*}-\sqrt{\ef}} \\
    &-&\frac{\sqrt{\hbar\omega_0}}{\sqrt{E_0-\Omega_0}}
    \left(\log\frac{\sqrt{E_0-\Omega_0}+\sqrt{E_{\rm F}}}
    {\sqrt{E_0-\Omega_0}-\sqrt{E_{\rm F}}}+i\pi\right) 
    \nonumber \\
    &+&\frac{1}{2}\sqrt{\frac{\hbar\omega_0}{E_{\rm F}}}
    \Bigg(
      i\pi\log\frac{\sqrt{E_{\rm F}-\Omega_0}+\sqrt{E_{\rm F}}}
      {\sqrt{E_{\rm F}-\Omega_0}-\sqrt{E_{\rm F}}} \nonumber \\
      &+&\text{Li}_2\frac{2\,\sqrt{E_{\rm F}}}
      {\sqrt{E_{\rm F}}+\sqrt{E_{\rm F}+\Omega_0^*}}
      +\text{Li}_2\frac{2\,\sqrt{E_{\rm F}}}
      {\sqrt{E_{\rm F}}-\sqrt{E_{\rm F}+\Omega_0^*}} \nonumber \\
      &-&\text{Li}_2\frac{2\,\sqrt{E_{\rm F}}}
      {\sqrt{E_{\rm F}}+\sqrt{E_{\rm F}-\Omega_0}}
      -\text{Li}_2\frac{2\,\sqrt{E_{\rm F}}}
      {\sqrt{E_{\rm F}}-\sqrt{E_{\rm F}-\Omega_0}}
    \Bigg)
  \Bigg],\nonumber 
  \end{eqnarray}
}
and
\nk{
  \begin{eqnarray}\label{eq.meff.dyson.dop}
  &&\frac{m^*}{m_0}=
  \Bigg[
    1+\frac{\alpha(\hbar\omega_0)^{3/2}}
    {2\pi\,\left(E_0+\Omega_0^*\right)^{3/2}}\\
    &&\times\left(
      \log\frac{\sqrt{E_0+\Omega_0^*}+\sqrt{E_{\rm F}}}
      {\sqrt{E_0+\Omega_0^*}-\sqrt{E_{\rm F}}}
      +\frac{2\sqrt{E_{\rm F}\,\left(E_0+\Omega_0^*\right)}}
      {E_0-E_{\rm F}+\Omega_0^*}
    \right)\nonumber\\
    &&-\frac{\alpha(\hbar\omega_0)^{3/2}}
    {2\pi\,\left(E_0-\Omega_0\right)^{3/2}}\nonumber\\
    &&\times\Bigg(
      \log\frac{\sqrt{E_0-\Omega_0}+\sqrt{E_{\rm F}}}
      {\sqrt{E_0-\Omega_0}-\sqrt{E_{\rm F}}}
      +\frac{2\sqrt{E_{\rm F}\left(E_0-\Omega_0\right)}}
      {E_0-E_{\rm F}-\Omega_0}
      -i\pi
    \Bigg)
  \Bigg]\nonumber\\
  &&\times\Bigg[
    1+\frac{2\alpha\,(\hbar\omega_0)^{3/2}}{3\pi}\nonumber \\
    &&\times\Bigg(
      \frac{E_0+\W0^*}{\ef\left(E_0-\ef+\W0^*\right)^2}
      -\frac{E_0-\W0}{\ef\left(E_0-\ef-\W0\right)^2} \nonumber\\
      &&+\frac{\tanh^{-1}\frac{\sqrt{E_{\rm F}}}{\sqrt{E_0+\Omega_0^*}}}
      {\left(E_0+\Omega_0^*\right)^{3/2}}
      -\frac{\tanh^{-1}\frac{\sqrt{E_{\rm F}}}{\sqrt{E_0-\Omega_0}}-\pi/2}
      {\left(E_0-\Omega_0\right)^{3/2}}\nonumber\\
      &&-\frac{\sqrt{E_{\rm F}}\left(E_{\rm F}-2E_0-2\Omega_0^*\right)}
      {\left(E_0\hspace{-0.8mm}+\hspace{-0.8mm}\Omega_0^*\right)
      \hspace{-0.8mm}
      \left(E_{\rm F}\hspace{-0.8mm}-\hspace{-0.8mm}E_0\hspace{-0.8mm}
      -\hspace{-0.8mm}\Omega_0^*\right)^2}
      \hspace{-0.5mm}+\hspace{-0.5mm}
      \frac{\sqrt{E_{\rm F}}\left(E_{\rm F}-2E_0+2\Omega_0\right)}
      {\left(E_0\hspace{-0.8mm}-\hspace{-0.8mm}\Omega_0\right)
      \hspace{-0.8mm}\left(E_{\rm F}\hspace{-0.8mm}
      -\hspace{-0.8mm}E_0\hspace{-0.8mm}+\hspace{-0.8mm}\Omega_0\right)^2}
    \Bigg)
  \Bigg]^{-1}.\nonumber
  \end{eqnarray}
}
As in the undoped case, all quantities in the Dyson approach are defined
self-consistently. In the cumulant approach, the QP energy and effective 
mass are again evaluated at the independent particle energy, and the
self-energy is always linear in $\alpha$. The $k$-dependent QP energy in 
the cumulant approach is given by:
  \begin{eqnarray}\label{eq.ek.cumul.dop}
  E_k^<&=&\epsilon_k\\
	  &+&\hspace{-0.8mm}\text{Re}\hspace{-0.5mm}\Bigg[
    \frac{\alpha\,(\hbar\omega_0)^{3/2}}{\pi\,\sqrt{\epsilon_k+\Omega_0}}
    \log\frac{\sqrt{\epsilon_k+\Omega_0}+\sqrt{E_{\rm F}}}
    {\sqrt{\epsilon_k+\Omega_0}-\sqrt{E_{\rm F}}}
    -\Sigma^<_{k_{\rm F}}(E_{\rm F})
  \Bigg].\nonumber
  \end{eqnarray}
At the $\Gamma$ point, this becomes: 
  \begin{eqnarray}\label{eq.en.cumul.dop}
  &&\frac{E_0^<}{\hbar\omega_0} = \frac{\alpha}{\pi}
  \text{Re}\Bigg[
    \log\frac{\sqrt{\Omega_0}+\sqrt{E_{\rm F}}}
    {\sqrt{\Omega_0}-\sqrt{E_{\rm F}}} 
    +\frac12\sqrt{\frac{\hbar\omega_0}{E_{\rm F}}} \\
    &&\times \Bigg(
      \text{Li}_2\frac{2\,\sqrt{E_{\rm F}}}
      {\sqrt{E_{\rm F}}+\sqrt{E_{\rm F}+\Omega_0}}
      +\text{Li}_2\frac{2\,\sqrt{E_{\rm F}}}
      {\sqrt{E_{\rm F}}-\sqrt{E_{\rm F}+\Omega_0}}
    \Bigg)
  \Bigg]. \nonumber
  \end{eqnarray}
The cumulant effective mass is given by
  \begin{eqnarray}\label{eq.meff.cumul.dop}
  \frac{m^*}{m_0}&=&
  \Bigg[
    1-\frac{\alpha}{2\pi}
    \left(
      \log\frac{\sqrt{\Omega_0}+\sqrt{E_{\rm F}}}
      {\sqrt{\Omega_0}-\sqrt{E_{\rm F}}}
      +\frac{2\sqrt{E_{\rm F}\,\Omega_0}}{\Omega_0-E_{\rm F}}
    \right)\nonumber\\
    &-&\frac{2\alpha}{3\pi\,\sqrt{E_{\rm F}}}
    \left(
      \sqrt{\hbar\omega_0}-\sqrt{E_{\rm F}}
      \tanh^{-1}\frac{\sqrt{E_F}}{\sqrt{\Omega_0}}
    \right)
  \Bigg]^{-1}\!\!.\hspace{0.3cm}
\end{eqnarray}
These expressions are illustrated in Fig.~\ref{fig5}. The dependence of the
QP renormalization on the coupling strength $\alpha$ and doping level 
$E_{\rm F}$ for the Dyson and cumulant approaches are shown in 
Figs.~\ref{fig5}(a) and (c), respectively. Figures~\ref{fig5}(b) and (d) 
show the corresponding effective masses.

The darker the shade of green in Figs.~\ref{fig5}(a) and (c), the lower the 
QP energy lies below the Fermi level.
This situation corresponds to a stable dressed electron, precisely as in the 
case of the undoped \F\ model. Conversely, the red areas in 
Figs.~\ref{fig5}(a) and (c) indicate that the QP peak would lie 
\textit{above} $E_{\rm F}$, causing a breakdown of the Fermi surface. This 
latter scenario is unphysical, and underscores the limitations of using a 
second-order electron-phonon self-energy. 

Moreover, moving to the effective masses shown in Figs.~\ref{fig5}(b) and
(d), we find areas in the phase space where overshooting mass 
renormalization leads to an inversion of the curvature in the QP spectrum,
and a negative effective mass (shown in red). We refer to this effect as 
'anomalous mass enhancement'.

\nk{
We emphasize that the data presented in Fig.~\ref{fig5} do not take into 
account free-carrier screening of the electron-phonon matrix elements. As 
we show in Sec.~\ref{sec.res.screened} and Fig.~\ref{fig9}, the inclusion 
of free-carrier screening extends the validity range of both the Dyson and 
cumulant approaches, but unphysical solutions still exist in the region 
$E_{\rm F}<\hbar\omega_0$, i.e. in the \textit{anti-adiabatic} regime. 
The insets in Figs.~\ref{fig5}(b) and (d) show an enlarged view of the
low-doping limit for small $\alpha$. Note that the color bar has been
extended with respect to the full image.}

In fact, this situation is reminiscent of \F's theory of superconductivity,\cite{Frohlich1950b} which 
incorrectly predicts an inversion of the band curvature at strong 
coupling.\cite{Kohn1951} This artifact was later resolved within the 
Bardeen-Cooper-Schrieffer theory, where the self-energy is evaluated 
self-consistently as opposed to perturbatively,\cite{Bardeen1957} and the 
resulting band structure features an energy gap, instead of inverted bands.

\nk{
The anomalous mass renormalization in the presence of doping, and the emergence of critical
values for $\alpha$ and $\ef$ is a consequence of an intricate dependence of $\Sigma$ on the
Fermi energy: In appendix~\ref{sec.app.mass}, we show that the \textsl{curvature} of the
(lesser) self-energy exhibits a singularity near $\ef=0$. The effective mass is defined as:
\begin{equation}\label{def.meff}
	m^*=\left[\frac{1}{m_0}
	+\frac{1}{\hbar^2}\frac{d^2\,\text{Re}\Sigma_k(E_k)}{dk^2}\right]^{-1}_{k=0},
\end{equation}
and $d^2\Sigma/dk^2<0$ for $\ef\ge0$. At large doping levels, the (negative) curvature of 
$\Sigma$ is small, and so is the mass enhancement due to Eq.~\eqref{def.meff}. With decreasing
$\ef$, the curvature $d^2\Sigma/dk^2$ approaches $-1/m_0$ from above, and the effective 
mass can reach arbitrarily large values. Beyond the critical value of $\ef$, as the magnitude of
$d^2\Sigma/dk^2$ keeps increasing, Eq.~\eqref{def.meff} becomes negative and the 
quasi-particle picture breaks down.

This behavior occurs in both the Dyson and cumulant aproaches, and is also independent of 
the specific implementation of the cumulant method: In appendix~\ref{sec.app.mass}, we 
show that the small-$E_{\rm F}$ behavior of the retarded cumulant\cite{Kas2014} is very 
similar.}

It is instructive to consider the QP energy and mass renormalization near this 
singularity in $\Sigma$. For small coupling strengths $\alpha\ll1$, we find the following 
expressions as we take the extreme anti-adiabatic limit $\ef\ll\hbar\omega_0$ in the 
Dyson approach:
  \begin{equation}\label{eq.en.dyson.ef}
  \frac{E_0}{\hbar\omega_0}
  =\frac{2\alpha}{\pi}\sqrt{\frac{\ef}{\w0}}
  +\frac{\alpha}{6}\frac{\ef}{\w0}
  +\mathcal{O}(E_{\rm F}/\hbar\omega_0)^{3/2},
  \end{equation}
  \begin{equation}\label{eq.meff.dyson.ef}
  \frac{m^*}{m_0}=
  -\frac{4\alpha}{3\pi}\sqrt{\frac{\w0}{\ef}}
  +\left(1+\frac{5\alpha}{6}\right)
  -\frac{8\alpha}{3\pi}\sqrt{\frac{\ef}{\w0}}
  +\mathcal{O}(E_{\rm F}/\hbar\omega_0)^{3/2}
  \end{equation}
while for the cumulant expansion we obtain:
\begin{equation}\label{eq.en.cumul.ef}
  \frac{E_0^<}{\hbar\omega_0}=
  \frac{\alpha}{\pi}\sqrt{\frac{\ef}{\w0}}
  +\mathcal{O}(E_{\rm F}/\hbar\omega_0)^{3/2},
\end{equation}
\begin{equation}\label{eq.meff.cumul.ef}
  \frac{m^*}{m_0}=\frac{2\alpha}{3\pi}\sqrt{\frac{\w0}{\ef}}
  +1+\frac{4\alpha}{3\pi}\sqrt{\frac{\ef}{\w0}}
  +\mathcal{O}(E_{\rm F}/\hbar\omega_0)^{3/2}.
\end{equation}
Note that to obtain Eqs.~\eqref{eq.en.dyson.ef}-\eqref{eq.meff.cumul.ef},
we have first taken the limit for small $\alpha$ followed by the limit for 
small $\ef$, i.e. we remain within the green area in Figs.~\ref{fig5}(b) 
and (d). 

Within the range of physical values for $(\alpha,\ef)$, we find that the 
Dyson approach tends to provide more stable solutions in the phase space, 
cf. Fig.~\ref{fig5}. Nevertheless, given that for the doped \F\ solid there 
exists no solution equivalent to Feynman's treatment of the \F\ 
polaron, it is difficult to judge whether the Dyson or cumulant approach 
yields the more accurate QP and mass renormalization.

For both approaches, we observe that combinations of intermediate-to-high 
coupling strengths and low doping levels quickly become problematic, while 
high doping levels and low coupling strengths lead to meaningful results.
Probably, this was to be expected, since the Migdal theorem (which 
underpins the Fan-Migdal self-energy) is only valid within the adiabatic 
approximation, $\hbar\omega_0 \ll E_{\rm F}$.\cite{Migdal1958}

This result is significant for the interpretation of experimental data. 
Commonly, the curvature of the QP band obtained from from ARPES 
measurements is used to determine the dressed mass and hence the coupling 
strength $\alpha$. In the literature, the coupling strength is often 
determined by using the formula for the undoped \F\ model, as given by 
Eqs.~\eqref{eq.mass.dyson.undop} and \eqref{eq.mass.cumul.undop}. However, 
the experimental setup for ARPES always requires a small but finite Fermi 
sea from which electrons can be excited. As discussed above, the physics
of such a system are likely better captured by 
Eqs.~\eqref{eq.en.dyson.ef}-\eqref{eq.meff.cumul.ef}.

Moving to the polaron satellites, we see from Figs.~\ref{fig4}(c) and (d)
that the Dyson approach again produces a single satellite starting with a 
broad area of low but non-zero spectral weight at one phonon energy below
the Fermi edge. Note that this area extends beyond the Fermi momentum
$k_{\rm F}$, suggesting that unoccupied electronic states could also 
emit a phonon upon excitation. This does not conform to experiment, and is 
an artefact of formulating the Fan-Migdal self-energy as a retarded 
quantity, i.e. treating occupied and unoccupied states within the same 
self-energy. At the lower end of the satellite structure, we recover a 
sharp peak whose intensity is almost of the order of the main QP peak.

On the other hand, the cumulant approach, shown in Figs.~\ref{fig4}(e) and 
(f), produces several satellites at exactly integer multiples of the boson 
energy, with peak intensities following a Poisson distribution.

Moreover, our analytical approach indicates that the satellites consist of
doublets: \nk{Consider Fig.~\ref{fig.new}, in which we juxtapose the imaginary self-energy
at $\epsilon_k=0.1\hbar\omega_0$, also shown in Fig.~\ref{fig4}(b), with the corresponding
satellite function which yields the double-peaked satellite of Fig.\ref{fig4}(e).
At finite doping, phonon emission and absorption processes can occur at all energies 
between $\epsilon_{k=0}$ and $\ef$, which causes the finite imaginary self-energy in the
energy range $\left[-\hbar\omega_0,-\hbar\omega_0+\ef\right]$. This is shown as the 
'phonon' line in blue in Fig.~\ref{fig.new}. Mathematically, this is due to the negative
argument of the logarithm in Eq.~\eqref{eq.dop.sigma.l} at the given values, which is 
independent of the electronic lifetime broadening $i\eta$. The energy of the lower edge of
the satellite, $\hbar\omega=\epsilon_{k}-\hbar\omega_0$, is equal to the singularity of 
the non-interacting electron Green's function shifted by the phonon energy. This is 
indicated by the green line 'electron' in Fig.~\ref{fig.new}. In numerical calculations, 
the peak height of this singularity is determined by the parameter $i\eta$, and it enters 
the self-energy through 
$L(\sqrt{E_{\rm F}/\epsilon_k},\sqrt{(\hbar\omega+\Omega_0^*)/\epsilon_k}$.
	
In some of the early work on the cumulant spectra of polarons, it was 
suspected that the satellite peak would simply follow the dispersion of the
QP band.\cite{Caruso2018} By deriving all involved quantities analytically,
we are able to uncover an even more nuanced picture of the cumulant 
satellites.}

Conversely, our analytical solutions exhibit a secondary peak whose 
dispersion is inverted with respect to the main satellite peak, leading to 
a near-elliptical feature. The energy separation of the two structures 
that constitute the satellite peak equals the Fermi energy. 
This non-trivial satellite structure may be related to the finite 
spectral weight between satellites observed in experiment, and it could 
provide an explanation for some of the broadening of spectral features seen
in ARPES but not yet reproduced in previous \textit{ab initio} 
calculations; cf. e.g. Refs.~\citenum{WangZ2016,Riley2018}.

\subsection{Finite Fermi level including free-carrier screening}
\label{sec.res.screened}

In this last section we consider the most complex scenario: a \F\ model with doping
as well as screening of the electron-phonon interaction by free carriers.
Due to the intricate dependence of the screened matrix element on the
wave vector $q$, see Eq.~\eqref{eq.def.gscreened}, we express the 
screened self-energy in terms of a one-dimensional integral in $q$. The
screened lesser self-energy is given by:
  \begin{eqnarray}\label{eq.scr.sigma.l}
  &&\Sigma^<_k(\omega) =
  -\frac{\alpha\,(\hbar\omega_0)^{3/2}}{2\pi\,\sqrt{\epsilon_k}} 
  \nonumber \\
  &&\times\Bigg[
    \int_0^{k_{\rm F}-k}\frac{dq}{|\varepsilon^\text{RPA}(q)|^2\,q}
    \log\frac{\hbar\omega-\frac{\hbar^2(k+q)^2}{2m_0}+\Omega_0}
    {\hbar\omega-\frac{\hbar^2(k-q)^2}{2m_0}+\Omega_0} \nonumber \\
  &&+\int_{k_{\rm F}-k}^{k_{\rm F}+k}
    \frac{dq}{|\varepsilon^\text{RPA}(q)|^2\,q}
    \log\frac{\hbar\omega-E_{\rm F}+\Omega_0}
    {\hbar\omega-\frac{\hbar^2(k-q)^2}{2m_0}+\Omega_0}
  \Bigg]\nonumber \\
  &&-\text{Re}\Sigma^<_{k_{\rm F}}(E_{\rm F}),
  \end{eqnarray}
with
  \begin{eqnarray}\label{eq.scr.sigma.l.kf}
  &&\Sigma^<_{k_{\rm F}}(E_{\rm F})=
  \frac{\alpha\,\hbar\omega_0}{2\pi}\sqrt{\frac{\hbar\omega_0}{E_{\rm F}}}
  \nonumber \\
  &&\times \int_0^{2k_{\rm F}}\frac{dq}{|\varepsilon^\text{RPA}(q)|^2\,q}
  \log\left(1-\frac{\hbar^2(q^2-2k_{\rm F}q)}{2m_0\,\Omega_0}\right).
  \end{eqnarray}
The screened greater self-energy is found to be:
  \begin{eqnarray}\label{eq.scr.sigma.g}
  &&\Sigma^>_k(\omega) =
  -\frac{\alpha\,(\hbar\omega_0)^{3/2}}{2\pi\,\sqrt{\epsilon_k}} 
  \nonumber \\
  &&\times\Bigg[
    \int_{k_{\rm F}-k}^{k_{\rm F}+k}
    \frac{dq}{|\varepsilon^\text{RPA}(q)|^2\,q}
    \log\frac{\hbar\omega-\frac{\hbar^2(k+q)^2}{2m_0}-\Omega_0}
    {\hbar\omega-E_{\rm F}-\Omega_0} \nonumber \\
    &&+\int_{k_{\rm F}+k}^\infty\frac{dq}{|\varepsilon^\text{RPA}(q)|^2\,q}
    \log\frac{\hbar\omega-\frac{\hbar^2(k+q)^2}{2m_0}-\Omega_0}
    {\hbar\omega-\frac{\hbar^2(k-q)^2}{2m_0}-\Omega_0}
  \Bigg] \nonumber \\
  &&-\text{Re}\Sigma^>_{k_{\rm F}}(E_{\rm F}),
  \end{eqnarray}
with
  \begin{eqnarray}
  &&\Sigma^>_{k_{\rm F}}(E_{\rm F}) =
  -\frac{\alpha\,\hbar\omega_0}{2\pi}\sqrt{\frac{\hbar\omega_0}{E_{\rm F}}} 
  \nonumber \\
  &&\times\Bigg[
    \int_0^{2k_{\rm F}}
    \frac{dq}{|\varepsilon^\text{RPA}(q)|^2\,q}
    \log\left(1+\frac{\hbar^2(q^2-2k_{\rm F}q)}{2m_0\,\Omega_0}\right)
    \nonumber \\
    &&+\int_{2k_{\rm F}}^\infty\frac{dq}{|\varepsilon^\text{RPA}(q)|^2\,q}
    \log\frac{1+\frac{\hbar^2(q^2+2k_{\rm F}q)}{2m_0\,\Omega_0}}
    {1+\frac{\hbar^2(q^2-2k_{\rm F}q)}{2m_0\,+\Omega_0}}
  \Bigg].
  \end{eqnarray}
Expressions for the QP energy and effective mass for the screened case 
are provided in App.~\ref{sec.renorm.screened}.

The magnitude of electronic screening effects depends both on the given
Fermi level $E_{\rm F}$, as well as material-specific parameters. In the 
Lindhard function in Eq.~\eqref{eq.lindhard}, system properties enter 
through the Wigner-Seitz radius $r_s$, which depends on the bare mass of 
the conduction electrons $m_0$ and the dielectric constant $\varepsilon$ 
of the undoped solid in the high-frequency limit. 

In Fig.~\ref{fig6}, we show the screening function 
$|\varepsilon^\text{RPA}(q,\omega_0)|^{-2}$ as a function of doping level
$n_0$ for two types of systems. This function quantifies the suppression
of the
electron-phonon matrix element $g$ by free-carrier screening, because
the \F\ matrix elements appears to the second power in the self-energy, and
the angular integration over the phonon wavevector introduces a phase-space factor $4\pi q^2$. Therefore
the impact of the screening is to modify a function that scales as $q^2|g|^2 \sim 1$ to 
a function that scales as $|\varepsilon^\text{RPA}|^{-2} q^2|g|^2 \sim |\varepsilon^\text{RPA}|^{-2}$. If 
$|\varepsilon^\text{RPA}|^{-2}=1$, there is no screening by free carriers;
for $|\varepsilon^\text{RPA}|^{-2}=0$, all electron-phonon coupling is 
completely suppressed. The dielectric function in Fig.~\ref{fig6}(a) 
corresponds to a dilute electron gas with a high electron mass and 
intermediate dielectric constant, as realized e.g. in cubic SrTiO$_3$. We 
observe that the ability of the free carriers to screen long-range 
(small-$q$) electron-phonon coupling gradually increases over a typical 
doping range for conducting oxides ($10^{18}$-$10^{21}$~cm$^{-3}$). At the 
highest doping level shown, polar interactions are almost completely 
suppressed, indicating that we have reached the metallic limit.

By contrast, the screening function in Fig.~\ref{fig6}(b) 
is that of a dense electron gas, as realized e.g. in GaAs. In the latter 
case, a very low electron mass means that the free carriers can screen
long-range electron-phonon interactions very effectively already at 
relatively low doping levels. In these systems, \F\ coupling only plays a 
secondary role, therefore in the remainder of this manuscript we focus on dilute electron
gases such as the one in Fig.~\ref{fig6}(a).

Figure~\ref{fig7} illustrates the effect of this screening function on 
the real and imaginary parts of the self-energy at a doping level of 
$E_{\rm F}/\hbar\omega_0=0.8$. The corresponding unscreened and screened 
Dyson spectral functions are shown in Figs.~\ref{fig8}(a)-(d), while the 
second-order cumulant spectral functions are shown in 
Figs.~\ref{fig8}(e)-(h). In particular, note that the Fermi surface in the 
unscreened second-order cumulant spectrum in Fig.~\ref{fig8}(e) is breaking
down due to (unphysically) strong renormalization effects at the given 
values of $\alpha$ and $E_{\rm F}$. We expect the effective electron-phonon 
interaction in a real system to be closer to the situation shown in 
Fig.~\ref{fig8}(g), in which the polar coupling is partially suppressed.

Given the $q$ dependence of the dielectric function,
the effect of electron screening is not uniform, but rather stronger up to
a scattering wave vector $q=k_{\rm F}$, and then weaker for states $k$ and 
$k'$ which are farther apart. 

The significant effect of free-carrier screening on the total \F\, coupling
strength in the system is also seen in Fig.~\ref{fig9}, where we 
show the QP energy and effective mass renormalization in the doped \F\,
solid using the screened matrix element. The comparison to the corresponding
Fig.~\ref{fig5} is telling: We find that the region showing the 
effective mass anomaly is reduced, and in general we observe
weaker renormalization for a wide range of coupling strengths $\alpha$ and
doping levels. For $E_{\rm F}/\hbar\omega_0>1.5$, the effective mass
in the second-order cumulant expansion returns to values close to the
non-interacting system. 

Turning to the satellites in Figs.~\ref{fig8}(a) and (c), we observe 
that the energy difference between QP and satellite in the screened
system is reduced by approximately 50\%. This result underscores the pathological 
dependence of the satellite energy on the coupling strength in Dyson's 
approach. As the energy of the phonon is the same with or without screening
(barring frequency renormalization effects that we did not consider in this
work), there is no reason to expect the satellite energy to change as a 
result of free-carrier screening.

By contrast, the lower peak in the satellite structure in the screened 
cumulant spectrum [Fig.~\ref{fig8}(g)] remains separated from the QP peak 
by one phonon energy. As discussed above, the energy span of the satellite 
in the cumulant approach matches the size of the Fermi energy in the system.
At a Fermi level of $E_{\rm F}/\hbar\omega_0$=0.8, we find substantial 
spectral weight in the region in between the quasi-particle and the 
satellite. This finding is in remarkable agreement with the raw ARPES data 
reported in Refs.~\onlinecite{WangZ2016,Riley2018} for $n$-doped SrTiO$_3$
and EuO, respectively.

\nk{
\section{Experimental QP weight}\label{sec.qp.weight}
Wang et al.\cite{WangZ2016} present a very careful analysis of polaron satellites in the two-dimensional electron gas in doped SrTiO$_3$. The lowest doping level considered in this study is
$n_\text{2D}=2.9\times10^{13}$cm$^{-2}$. We can assume that the conduction band near $\Gamma$ is well described by three degenerate parabolic bands with an average band mass of $m_0=0.9$.\cite{Devreese2010} 
The considered doping level then corresponds to a Fermi wavevector of $k_{\rm F}$=0.041$a_0^{-1}$ and a Fermi energy of $E_{\rm F}=25$meV. With a phonon energy of $\hbar\omega_0=$100meV 
and a reported \F\ coupling strength of $\alpha\approx2-3$, they observe intermediate quasi-particle renormalization with $Z\approx0.2$.\\
\newline
Equations~\eqref{eq.dm.zk} and \eqref{eq.ce.zk} give the expressions for the Dyson and cumulant 
quasi-particle weights, respectively. To be able to evaluate them analytically, we have derived expressions for the
frequency derivative of the self-energy for finite Fermi energy. For the Dyson equation approach, we need the
derivative of the full retarded self-energy:
  \begin{eqnarray}\label{dw.sigma.r}
  &&\frac{1}{\hbar}\frac{\partial\Sigma_{k=0}(\omega)}{\partial \omega}
  =-\frac{\alpha\,(\hbar\omega_0)^{3/2}}
  {2\pi\,\left(\hbar\omega+\Omega_0^*\right)^{3/2}} \\
  &\times&\left(
    \log\frac{\sqrt{\hbar\omega+\Omega_0^*}+\sqrt{E_{\rm F}}}
    {\sqrt{\hbar\omega+\Omega_0^*}-\sqrt{E_{\rm F}}}
    +\frac{2\sqrt{E_{\rm F}\,\left(\hbar\omega+\Omega_0^*\right)}}
    {\hbar\omega+\Omega_0-E_{\rm F}}
  \right)\nonumber \\
  &+&\frac{\alpha\,(\hbar\omega_0)^{3/2}}
  {2\pi\,\left(\hbar\omega-\Omega_0\right)^{3/2}} \nonumber \\
  &\times&\left(
  \log\frac{\sqrt{\hbar\omega-\Omega_0}+\sqrt{E_{\rm F}}}
  {\sqrt{\hbar\omega-\Omega_0}-\sqrt{E_{\rm F}}}
  +\frac{2\sqrt{E_{\rm F}\,\left(\hbar\omega-\Omega_0\right)}}
  {\hbar\omega-\Omega_0-E_{\rm F}}+i\pi
  \right).\nonumber
  \end{eqnarray}
To obtain the Dyson quasi-particle weight at $k=0$, we evaluate this 
expression at $\hbar\omega=\epsilon_{k=0}+Z_{k=0}\epsilon_{k=0}$.

For the cumulant expansion, we find for the derivative:
  \begin{multline}\label{dw.sigma.l}
  \frac{1}{\hbar}\frac{\partial\Sigma^<_{k=0}(\omega=0)}{\partial \omega}\\
  =-\frac{\alpha}{2\pi}
  \left(
    \log\frac{\sqrt{\Omega_0}+\sqrt{E_{\rm F}}}
    {\sqrt{\Omega_0}-\sqrt{E_{\rm F}}}
    +\frac{2\sqrt{E_{\rm F}\,\Omega_0}}{\Omega_0-E_{\rm F}}
  \right).
  \end{multline}
We insert Eqs.~\eqref{dw.sigma.r} and \eqref{dw.sigma.l} into 
Eqs.~\eqref{eq.dm.zk} and \eqref{eq.ce.zk}, and evaluate the quasi-particle 
weight using the parameters
  \begin{equation}
  \alpha=3,\quad\hbar\omega_0=100\text{meV}\quad m_0=0.9\text{m}_e
  \quad E_{\rm F}=25\text{meV}.
  \nonumber
  \end{equation}
The resulting quasi-particle weight is reported in table~\ref{tab2}. We 
find that the calculated Dyson and cumulant QP weights over-estimate the 
measured weight by a factor of 2 and 1.5, respectively. For case of the 
cumulant expansion, this inconsistency between theoretical and experimental 
spectra has been studied in great detail in the literature, see e.g. 
Ref.~\citenum{Zhou2020}. In this study, the authors are able to show how 
additional satellite intensity arises from inelastic scattering of the 
outgoing photoelectron in experiment. 
}

\section{Conclusions}\label{sec.conc}

We have presented the doped Fr{\"o}hlich solid as a generalization of the 
Fr{\"o}hlich polaron problem to study the single-particle excitation 
spectra of doped polar oxides, as measured by ARPES experiments. To reach 
reliable conclusions that are not affected by numerical sampling of the 
electron-phonon scattering, we derived exact analytical expressions for the
electron self-energy in the presence of free carriers. These expressions 
allow to analyze in detail the role of Pauli blocking and free carrier 
screening in the electron spectral functions.

Our analytical approach has provided new insight into the \F\ polaron 
problem, and allows us to draw the following conclusions: 
To capture the low-energy many-body physics of doped polar semiconductors, 
especially in the presence of doping (as needed in ARPES measurements), it 
is crucial to explicitly account for the small but non-zero electron 
occupations in the conduction band. We have demonstrated that neglecting
finite occupations leads to incorrect satellite energetics and excessive 
electron-phonon renormalization. In the case of high doping levels, further
many-body effects in the form of free-carrier screening of the 
electron-phonon matrix element must also be included to achieve a 
meaningful description of QP shifts and effective mass renormalization.

We have derived analytical expressions for the renormalized band energy 
and effective mass of the doped \F\ model, and investigated the dependence 
of these quantities on the coupling strength and the doping level. We have 
found that a significant portion of the coupling-doping phase diagram 
exhibits regions with anomalously strong mass enhancement, as well as 
regions where the band curvature is inverted, leading to a breakdown of the
QP picture. These findings indicate that caution must be used when studying
electron-phonon coupling in doped polar materials, as the standard 
second-order Fan-Migdal self-energy might not provide a physically
accurate picture in the anomalous regions of the phase diagram identified 
in this work. 

We also found that, in the presence of doping, the mass renormalization 
depends both on the electron-phonon coupling strength \textit{and} the 
Fermi level. This finding implies that the use of the standard relation 
$m^*/m_0 = 1+\alpha/6$ for extracting the \F\ coupling from experiments is 
not justified, and should be replaced by the generalized expression 
obtained in this work, Eq.~(51).

In line with previous literature, we found that the second-order cumulant 
spectral function improves the description of polaron satellites compared 
to the conventional first-order Dyson approach. The intensity and binding 
energy of the sidebands provided by the cumulant method are in line with 
the equidistant satellites observed in experiments. On the other hand, the 
cumulant method appears to provide a worse description of the QP band, as 
compared to the Dyson approach. Indeed, we have shown that the cumulant 
approach leads to an inversion of the band curvature over a much wider 
region of the phase diagram as compared to the Dyson method, and that the
cumulant spectral function exhibits unphysical vertical streaks that are 
intrinsic to the theoretical framework (rather than being numerical 
artifacts). Our comparative analysis of the cumulant and Dyson approaches 
leads to suggest that a complete description of the \F\ problem in the 
presence of doping might require the inclusion of self-energy diagrams 
beyond the second order. In the meantime, we recommend that both approaches
be tested in future calculations, keeping in mind that the cumulant method 
appears more suited to describing satellites, and the Dyson method 
appears to describe QP bands better.

We hope that this work will stimulate further discussion on the role of 
doping in the electron-phonon interaction in polar insulators and 
semiconductors, and inspire additional investigations of the reliability 
and scope of the cumulant method and the Dyson approach in the study of 
electron-phonon effects in these materials.

\section*{Acknowledgements}

This work was supported by the Computational Materials Sciences Program 
funded by the U.S. Department of Energy, Office of Science, Basic Energy 
Sciences, under Award DE-SC0020129. The authors acknowledge the Texas 
Advanced Computing Center (TACC) at The University of Texas at Austin for 
providing HPC resources via the Frontera LRAC project DMR21002, and 
the U.S. Department of Energy Office of Science User Facility at Lawrence
Berkeley National Laboratory for providing resources via the National Energy 
Research Scientific Computing Center, NERSC award ERCAP0016747.

\appendix
\section{Debye-Waller self-energy}\label{sec.debye}
In this appendix, we show that the QP shift arising from the Debye-Waller 
self-energy\cite{Giustino2017} vanishes in the \F\ model.
We start from the compact expression for the Debye-Waller matrix element
\nk{in the rigid-ion approximation} derived in Ref.\citenum{Lihm2020}:
\begin{equation}\label{eq:debye}
	D^{\kappa \alpha \alpha'}(\textbf{k})
	=i\langle u_\textbf{k}|
	\big[\partial_{\Gamma\kappa\alpha}\hat{V}^\text{L},
	\hat{p}_{\alpha'}\big]|u_\textbf{k}\rangle,
\end{equation}
where $\alpha$, $\alpha'$ are Cartesian coordinates, $\kappa$ is the atomic
index, $u_\textbf{k}$ are the Bloch-periodic components of the electron 
wavefunctions, $\hat{V}^\text{L}$ is the long-range part of the interaction
potential, and 
\begin{equation}
	\hat{p}_\alpha=\sum_{\kappa\alpha'}Z_{\kappa\alpha\alpha'}
	\Delta\tau_{\kappa\alpha'}
\end{equation}
is the dipole moment in direction $\alpha$ arising from the displacement of
atom $\kappa$ with Born effective charge $Z_{\kappa\alpha\alpha'}$ along 
the direction $\Delta\tau_{\kappa\alpha'}$. The potential derivative in
Eq.\eqref{eq:debye} is defined as\cite{Giustino2017}
\begin{equation}
	\partial_{\textbf{q}\kappa\alpha}\hat{V}=
	\sum_p e^{-i\textbf{q}\cdot (\textbf{r}-\textbf{R}_p)}
	\frac{\partial V}{\partial\tau_{\kappa\alpha}}
	\Bigg|_{\textbf{r}-\textbf{R}_p},
\end{equation}
where $\textbf{R}_p$ is the lattice vector of the $p$-th unit cell in the
supercell, and $\tau_{\kappa\alpha}$ is the coordinate of atom $\kappa$
in cartesian direction $\alpha$.

In the doped Fr{\"o}hlich solid, the electronic states are planewaves
and the long-range part of the \F\, potential is
\cite{Verdi2015}
\begin{eqnarray}
	\hat{V}^\text{L}(\textbf{r})
	&=&-i\frac{4\pi}{4\pi\epsilon_0\,N\Omega_\text{UC}}
	\\
	&\times&\sum_{\substack{\textbf{q}\\\textbf{G}\ne-\textbf{q}}}
        \sum_{\kappa\alpha\alpha'}
        Z_{\kappa\alpha\alpha'}\Delta\tau_{\kappa\alpha'}
        \frac{(\textbf{q}+\textbf{G})_\alpha\,
	e^{i(\textbf{q}+\textbf{G})\cdot\textbf{r}}}
        {(\textbf{q}+\textbf{G})
        \cdot\pmb{\varepsilon}\cdot(\textbf{q}+\textbf{G})}.\nonumber
\end{eqnarray}
Here, $\textbf{G}$ is a reciprocal lattice vector, and $\pmb{\varepsilon}$
is the static dielectric tensor of the crystal. Using the isotropy of our
system to simplify the Born effective charge and dielectric tensors,
$Z_{\kappa\alpha\alpha'}=Z_\kappa\delta_{\alpha\alpha'}$ and
$\varepsilon_{\alpha\alpha'}=\epsilon_\infty\delta_{\alpha\alpha'}$,
we can calculate the derivative
\begin{equation}
	\partial_{\Gamma\kappa\alpha}
	=-\frac{i}{\epsilon_0\,\epsilon_\infty\,\Omega_\text{UC}}
	\sum_{\textbf{G}\ne0}\sum_\kappa Z_\kappa
	\frac{G_\alpha\,e^{i\textbf{G}\cdot\textbf{r}}}{|\textbf{G}|^2}.
\end{equation}
After calculating the commutator of this function with the dipole 
$\hat{p}_\alpha$, we obtain a lattice periodic expression, and as the 
eigenstates of our system are plane waves, we find
\begin{equation}
	\langle u_\textbf{k}|e^{i\textbf{G}\cdot\textbf{r}}|
	u_\textbf{k}\rangle = 0.
\end{equation}
This result shows that the Debye-Waller correction vanishes for the \F\ model.

\section{Derivation of the Fan-Migdal self-energy}\label{app.steps.undoped}

In this appendix, we outline the derivation of the various
self-energy expressions used in this manuscript, starting from the 
definition of the Fan-Migdal self-energy given in Eqs.~\eqref{eq.sigma.l} 
and \eqref{eq.sigma.g}.

\subsection{Single electron in the conduction band}

In the case of a single electron added to the conduction band,
the occupation factor
$f_{\textbf{k}+\textbf{q}}$ vanishes everywhere, cancelling all
contributions from the lesser self-energy. 

At the $\Gamma$ point ($\textbf{k}=0$), we integrate Eq.~\eqref{eq.sigma.g}
by introducing spherical coordinates. We write:
  \begin{equation}\label{eq.app.1}
  \Sigma^>_{k=0}(\omega)=4\pi\int_0^\infty\frac{dq\,q^2}{(2\pi)^3}
  \frac{|g(q)|^2}{\hbar\omega-\frac{\hbar^2q^2}{2m_0}-\Omega_0},
  \end{equation}
  and find:
  \begin{equation}
  \Sigma^>_{k=0}(\omega)=-i\,\frac{\alpha\,(\hbar\omega_0)^{3/2}}
  {\sqrt{\hbar\omega-\Omega_0}},
  \end{equation}
where we used the definition of the \F\, matrix element, 
Eq.~\eqref{eq.def.g2}. Equation~\eqref{eq.app.1} leads 
directly to Eq.~\eqref{eq.en.dyson.undop}.

For general $k$, we write Eq.~\eqref{eq.sigma.g} as 
  \begin{equation}
  \Sigma^>_\textbf{k}(\omega)=\int\frac{d\textbf{q}}{(2\pi)^3}
  \frac{|g(\textbf{q}-\textbf{k})|^2}
  {\hbar\omega-\frac{\hbar^2\textbf{q}^2}{2m_0}-\Omega_0}.
  \end{equation}
After transforming to spherical coordinates, we use the identity 
  \begin{equation}
  \int_0^\pi\frac{d\theta\,\sin\theta}{k^2+q^2-2kq\cos\theta}
  =\frac{1}{2kq}\log\frac{(k+q)^2}{(k-q)^2},
  \end{equation}
and the definition of the \F\, matrix element, Eq.~\eqref{eq.def.g2}, to obtain:
  \begin{eqnarray}\label{eq.app.2}
  \Sigma^>_k(\omega)=\frac{\alpha\,\hbar(\hbar\omega_0)^{3/2}}
  {2\pi\,k\,\sqrt{2m_0}}\int_0^\infty
  \frac{dq\,q}{\hbar\omega-\frac{\hbar^2q^2}{2m_0}-\Omega_0}
  \log\frac{(k+q)^2}{(k-q)^2}.\nonumber\\
  \end{eqnarray}
The integrand in Eq.~\eqref{eq.app.2} has the primitive:
  \begin{eqnarray}\label{eq.app.3}
  I(q)&=&-\Bigg[
    \log\frac{\hbar\omega-\frac{\hbar^2q^2}{2m_0}-\Omega_0}
    {\hbar\omega-\frac{\hbar^2k^2}{2m_0}-\Omega_0}
    \log\Bigg|\frac{q+k}{q-k}\Bigg| \\
    &+&\text{Li}_2\frac{\frac{\hbar(k+q)}{\sqrt{2m_0}}}
    {\frac{\hbar k}{\sqrt{2m_0}}-\sqrt{\hbar\omega-\Omega_0}}
    +\text{Li}_2\frac{\frac{\hbar(k+q)}{\sqrt{2m_0}}}
    {\frac{\hbar k}{\sqrt{2m_0}}+\sqrt{\hbar\omega-\Omega_0}} \nonumber \\
    &-&\text{Li}_2\frac{\frac{\hbar(k-q)}{\sqrt{2m_0}}}
    {\frac{\hbar k}{\sqrt{2m_0}}+\sqrt{\hbar\omega-\Omega_0}}
    -\text{Li}_2\frac{\frac{\hbar(k-q)}{\sqrt{2m_0}}}
    {\frac{\hbar k}{\sqrt{2m_0}}-\sqrt{\hbar\omega-\Omega_0}}
  \Bigg].\nonumber\hspace{3mm}
  \end{eqnarray}
We find $I(q=0)=0$; in the limit $q\rightarrow \infty$, only the 
dilogarithms $\text{Li}_2$ survive and converge to:
  \begin{equation}\label{eq.app.4}
  \lim_{q\rightarrow \infty}\,I(q)
  =-i\pi\log\frac{\sqrt{\hbar\omega-\Omega_0}+\sqrt{\epsilon_k}}
  {\sqrt{\hbar\omega-\Omega_0}-\sqrt{\epsilon_k}}.
  \end{equation}
Inserting Eq.~\eqref{eq.app.4} into Eq.~\eqref{eq.app.2}, we find 
Eq.~\eqref{eq.sigma.g.undoped} from the main text.

\subsection{Finite Fermi level in the conduction band}

We now consider the scenario where we have a finite electron density in the conduction band.
At $k=0$, the lesser self-energy is defined as
  \begin{equation}
  \Sigma^<_{k=0}(\omega) =\, 
  4\pi\int_0^{k_{\rm F}}\frac{dq\,q^2}{(2\pi)^3}
  \frac{|g(q)|^2}{\hbar\omega-\frac{\hbar^2q^2}{2m_0}+\Omega_0},
  \end{equation}
which can be integrated to give
  \begin{equation}
  \Sigma^<_{k=0}(\omega)
  =\frac{\alpha\,(\hbar\omega_0)^{3/2}}{\pi\,\sqrt{\hbar\omega+\Omega_0}}
  \log\frac{\sqrt{\hbar\omega+\Omega_0}+\sqrt{E_{\rm F}}}
  {\sqrt{\hbar\omega+\Omega_0}-\sqrt{E_{\rm F}}}.
  \end{equation}
For the greater self-energy, we have
  \begin{equation}
  \Sigma^>_{k=0}(\omega)=4\pi\int_{k_{\rm F}}^\infty
  \frac{dq\,q^2}{(2\pi)^3}
  \frac{|g(q)|^2}{\hbar\omega-\frac{\hbar^2q^2}{2m_0}-\Omega_0},
  \end{equation}
and hence
  \begin{eqnarray}
  \Sigma^>_{k=0}(\omega)=
  -\frac{\alpha\,(\hbar\omega_0)^{3/2}}{\pi\,\sqrt{\hbar\omega-\Omega_0}}
  \Bigg[
	  \log\frac{\sqrt{\hbar\omega-\Omega_0}+\sqrt{E_{\rm F}}}
	  {\sqrt{\hbar\omega-\Omega_0}-\sqrt{E_{\rm F}}}
    +i\pi
  \Bigg].\nonumber\\
  \end{eqnarray}
For general $k$, the lesser self-energy is obtained from an expression
similar to that of Eq.~\eqref{eq.app.3}, but with the sign of $\Omega_0$
inverted, which then needs to be evaluated at $q=0$ and at $q=k_{\rm F}$, respectively, to
give Eq.~\eqref{eq.dop.sigma.l}. The greater self-energy at $k\ne 0$ is 
obtained from Eq.~\eqref{eq.app.3} evaluated at $q=k_{\rm F}$ and 
$q\rightarrow \infty$, which leads to Eq.~\eqref{eq.dop.sigma.g}.

\subsection{Finite Fermi level including free-carrier screening}

In this section we provide details on the calculation of the self-energy in the presence
of free-carrier screening.
Inserting the screened coupling matrix element given in 
Eq.~\eqref{eq.sigma.g} into Eqs.~\eqref{eq.sigma.l} and \eqref{eq.sigma.g},
we obtain for the self-energy at $k=0$:
  \begin{equation}\label{eq.app.5}
  \Sigma^<_{k=0}(\omega)=\int_0^\infty \frac{dq\,q^2}{2\pi^2}
  \frac{|g(q)|^2}{|\varepsilon^\text{RPA}(q)|^2}
  \frac{\theta\left(k_{\rm F}^2-q^2\right)}
  {\hbar\omega-\frac{\hbar^2q^2}{2m_0}+\Omega_0},
  \end{equation}
  \begin{equation}\label{eq.app.6}
  \Sigma^>_{k=0}(\omega)=\int_0^\infty \frac{dq\,q^2}{2\pi^2}
  \frac{|g(q)|^2}{|\varepsilon^\text{RPA}(q)|^2}
  \frac{\theta\left(q^2-k_{\rm F}^2\right)}
  {\hbar\omega-\frac{\hbar^2q^2}{2m_0}-\Omega_0}.
  \end{equation}
Using the \F\, matrix element in Eq.~\eqref{eq.def.g2}, we recover 
Eqs.~\eqref{eq.scr.sigma.l} and \eqref{eq.scr.sigma.g} from the main text.

For general $k$, we write 
  \begin{eqnarray}
  \Sigma^<_k(\omega)&=&\int_0^{2\pi}d\phi \int_0^\pi d\theta \sin\theta 
  \int_0^\infty \frac{dq\,q^2}{(2\pi)^3}
  \frac{|g(q)|^2}{|\varepsilon^\text{RPA}(q)|^2} \nonumber \\
  &\times&  \frac{\theta\left(k_{\rm F}^2-(k^2+q^2+2kq\cos\theta)\right)}
  {\hbar\omega-\frac{\hbar^2(k^2+q^2+2kq\cos\theta)}{2m_0}+\Omega_0},
  \end{eqnarray}
  \begin{eqnarray}
  \Sigma^>_k(\omega)&=&\int_0^{2\pi}d\phi \int_0^\pi d\theta \sin\theta 
  \int_0^\infty \frac{dq\,q^2}{(2\pi)^3}
  \frac{|g(q)|^2}{|\varepsilon^\text{RPA}(q)|^2} \nonumber \\
  &\times&  \frac{\theta\left((k^2+q^2+2kq\cos\theta)-k_{\rm F}\right)}
  {\hbar\omega-\frac{\hbar^2(k^2+q^2+2kq\cos\theta)}{2m_0}-\Omega_0},
  \end{eqnarray}
where $\phi$ and $\theta$ are relative angles between vectors $\textbf{k}$ 
and $\textbf{q}$. For the integration over angle $\theta$, we use the 
substitution
$x=k_{\rm F}-k^2-q^2-2kq\cos\theta$ to obtain:
  \begin{eqnarray}\label{eq.app.7}
  \Sigma^<_k(\omega)&=&\frac{2\pi}{\hbar}
  \frac{1}{2k}  \int_0^\infty \frac{dq\,q}{(2\pi)^3}
  \frac{|g(q)|^2}{|\varepsilon^\text{RPA}(q)|^2} \nonumber \\
  &\times&  \int_{k_{\rm F}^2-(k+q)^2}^{k_{\rm F}^2-(k-q)^2} dx 
  \frac{\theta\left(x\right)}
  {\hbar\omega-\frac{\hbar^2(k_{\rm F}^2-x)}{2m_0}+\Omega_0},
  \end{eqnarray}
  \begin{eqnarray}\label{eq.app.8}
  \Sigma^>_k(\omega)&=&\frac{2\pi}{\hbar}
  \frac{1}{2k}  \int_0^\infty \frac{dq\,q}{(2\pi)^3}
  \frac{|g(q)|^2}{|\varepsilon^\text{RPA}(q)|^2} \nonumber \\
  &\times&  \int_{k_{\rm F}^2-(k+q)^2}^{k_{\rm F}^2-(k-q)^2} dx 
  \frac{\theta\left(-x\right)}
  {\hbar\omega-\frac{\hbar^2(k_{\rm F}^2-x)}{2m_0}-\Omega_0}.
  \end{eqnarray}
As we are using spherical coordinates, we have \\
$k\ge0$, $q\ge0$ and $k_{\rm F}>0$, and hence 
  \begin{equation}
  k_{\rm F}^2-(k-q)^2\ge k_{\rm F}^2-(k+q)^2,
  \end{equation}
allowing us to identify three ranges for the integration over $q$: 
For the lesser self-energy [Eq.~\eqref{eq.app.7}] we have: (i)
If $k_{\rm F}^2\ge(k+q)^2$, then $\theta\left(x\right)= 1$ and
  \begin{multline}\label{eq.app.9}
  \int_{k_{\rm F}^2-(k+q)^2}^{k_{\rm F}^2-(k-q)^2}
  \frac{dx}{\hbar\omega-\frac{\hbar^2(k_{\rm F}^2-x)}{2m_0}+\Omega_0}
  \\
  =\log\frac{\hbar\omega-\frac{\hbar^2(k+q)^2}{2m_0}+\Omega_0}
  {\hbar\omega-\frac{\hbar^2(k-q)^2}{2m_0}+\Omega_0}.
  \end{multline}
(ii) If $(k+q)^2\ge k_{\rm F}^2\ge(k-q)^2$, we have
  \begin{multline}\label{eq.app.10}
  \int_{0}^{k_{\rm F}^2-(k-q)^2}
  \frac{dx}{\hbar\omega-\frac{\hbar^2(k_{\rm F}^2-x)}{2m_0}+\Omega_0}\\ 
  =\log\frac{\hbar\omega-E_{\rm F}+\Omega_0}
  {\hbar\omega-\frac{\hbar^2(k-q)^2}{2m_0}+\Omega_0}.
  \end{multline}
(iii) For $(k-q)^2\ge k_{\rm F}$, the Heaviside function in 
\eqref{eq.app.7} vanishes everywhere. In particular, there is no 
contribution for $q\ge k_{\rm F}$. Combining Eqs.~\eqref{eq.app.7}, 
\eqref{eq.app.9} and \eqref{eq.app.10}, we recover 
Eq.~\eqref{eq.scr.sigma.l} from the main text.

For the greater self-energy [Eq.~\eqref{eq.app.8}], we have (i)
$\theta\left(-x\right)=0$ whenever $k_{\rm F}^2\ge(k+q)^2$, 
cancelling all contributions for $q<k_{\rm F}$. (ii) If 
$(k+q)^2\ge k_{\rm F}^2\ge(k-q)^2$, we have
  \begin{multline}\label{eq.app.11}
  \int_{k_{\rm F}^2-(k+q)^2}^{0}
  \frac{dx}{\hbar\omega-\frac{\hbar^2(k_{\rm F}^2-x)}{2m_0}-\Omega_0}\\ 
  =\log\frac{\hbar\omega-\frac{\hbar^2(k+q)^2}{2m_0}-\Omega_0}
  {\hbar\omega-\frac{\hbar^2(k-q)^2}{2m_0}-\Omega_0}.
  \end{multline}
(iii) For $(k-q)^2\ge k_{\rm F}^2$, we obtain
  \begin{multline}\label{eq.app.12}
  \int_{k_{\rm F}^2-(k+q)^2}^{k_{\rm F}^2-(k-q)^2} 
  \frac{dx}{\hbar\omega-\frac{\hbar^2(k_{\rm F}^2-x)}{2m_0}-\Omega_0} \\
  =\log\frac{\hbar\omega-\frac{\hbar^2(k+q)^2}{2m_0}-\Omega_0}
  {\hbar\omega-\frac{\hbar^2(k-q)^2}{2m_0}-\Omega_0}.
  \end{multline}
Combining Eqs.~\eqref{eq.app.8}, \eqref{eq.app.11} and \eqref{eq.app.12}, 
we recover Eq.~\eqref{eq.scr.sigma.g} from the main text.

\section{Derivation of the effective mass}\label{sec.app.mass}
\subsection{Dyson effective mass}
The effective mass $m^*$ corresponding to the QP energy in Dyson's approach,
\begin{equation}
	E_k=\epsilon_k+\text{Re}\,\Sigma_k(E_k),
\end{equation}
is given at $k=0$ by
  \begin{equation}
  \frac{1}{m^*}=\frac{1}{\hbar^2}\frac{d^2E_k}{dk^2}\Bigg|_{k=0}
  =\frac{1}{m_0}+\frac{1}{\hbar^2}\frac{d^2\,\text{Re}\Sigma_k(E_k)}{dk^2}
  \Bigg|_{k=0}.
\end{equation}
We can express the momentum dependence of the self-energy in terms of the
associated bare electron energy $\epsilon_k$, and write:
\begin{equation}
	\frac{d^2\Sigma(\epsilon_k,E_k)}{dk^2}
	=\frac{\partial\Sigma(\epsilon_k,E_k)}
	{\partial\epsilon_k}\frac{d^2\,\epsilon_k}{dk^2}
	+\frac{\partial\Sigma(\epsilon_k,E_k)}
	{\partial E_k}\frac{d^2 E_k}{dk^2},
\end{equation}
and hence\cite{Schlipf2018}
\begin{equation}
	\frac{m^*}{m_0}
	=\frac{1-\partial\text{Re}\Sigma(\epsilon_k,E_k)/\partial E_k}
	{1+\partial\text{Re}\Sigma(\epsilon_k,E_k)/\partial \epsilon_k}
	\Bigg|_{k=0}.
\end{equation}
\subsection{Cumulant effective mass}
As the QP energy in the cumulant expansion is simply given by
  \begin{equation}
  E^<_k=\epsilon_k+\text{Re}\,\Sigma^<_k(\epsilon_k),
  \end{equation}
the corresponding effective mass is equal to
  \begin{equation}
  \frac{m^*}{m_0}=
  \left[1+\frac{1}{\hbar}
    \frac{\partial\,\text{Re}\,\Sigma^<_k(\omega)}{\partial \omega}
    +\frac{\partial\,\text{Re}\,\Sigma^<_k(\omega)}{\partial \epsilon_k}
  \right]^{-1}_{\epsilon_k=0,\omega=0}.
  \end{equation}
As the cumulant self-energy is linear in $\alpha$, the small-$\alpha$ 
expansion of the effective mass becomes
  \begin{equation}
  \frac{m^*}{m_0}=1-\frac{1}{\hbar}
  \frac{\partial\,\text{Re}\,\Sigma^<_k(\omega)}{\partial \omega}
  -\frac{\partial\,\text{Re}\,\Sigma^<_k(\omega)}{\partial \epsilon_k}
  +\mathcal{O}(\alpha^2).
  \end{equation}
\nk{
\subsection{Singularity in $d^2\Sigma/dk^2$}
We can calculate the curvature of $\Sigma$ at $k=0$ for the unscreened system by starting from
Eq.~\eqref{eq.scr.sigma.l} and setting $\varepsilon^\text{RPA}\equiv 1$. After taking the derivative
we are left with the integral in $q$:
\begin{eqnarray}\label{step}
	\frac{\partial\,\Sigma^<_k(\omega)}{\partial\epsilon_k}\Bigg|_{k=0}
	&=&-\frac{2\alpha\,\hbar\omega_0\,\sqrt{\ef\,\hbar\omega_0}}{3\pi} \\
	&\times&\int_0^{k_{\rm F}}\frac{dq}{k_{\rm F}}
	\frac{\frac{\hbar^2q^2}{2m_0}+3\hbar\omega+3\Omega_0}
	{\left(\frac{\hbar^2q^2}{2m_0}-\hbar\omega-\Omega_0\right)^3}\nonumber \\
	&-&\frac{2\alpha\,\hbar\omega_0}{3\pi}
	\sqrt{\frac{\hbar\omega_0}{\ef}}
	\frac{\hbar\omega+\Omega_0}{\left(\hbar\omega-\ef+\Omega_0\right)^2}\nonumber
\end{eqnarray}
From the second term in Eq.~\eqref{step}, its divergent behavior at 
$\ef\rightarrow0$ is apparent. After evaluation of the integral, we find
  \begin{eqnarray}
  \frac{\partial\,\Sigma^<_k(\omega)}{\partial\epsilon_k}\Bigg|_{k=0}
  &=&-\frac{2\alpha\,\hbar\omega_0\,\sqrt{\ef\,\hbar\omega_0}}{3\pi} \\
  &\times&\Bigg[
    \frac{\ef-2\hbar\omega-2\Omega_0}
    {\left(\hbar\omega+\Omega_0\right)
    \left(\ef-\hbar\omega-\Omega_0\right)^2}\nonumber \\
    &-&\frac{1}{\sqrt{\ef}\left(\hbar\omega+\Omega_0\right)^{3/2}}
    \tanh^{-1}\frac{\sqrt{\ef}}{\sqrt{\hbar\omega+\Omega_0}}
  \Bigg]\nonumber \\
  &-&\frac{2\alpha\,\hbar\omega_0}{3\pi}
  \sqrt{\frac{\hbar\omega_0}{\ef}}
  \frac{\hbar\omega+\Omega_0}{\left(\hbar\omega-\ef+\Omega_0\right)^2}.
  \nonumber
  \end{eqnarray}
}
\subsection{Effective mass of the retarded cumulant}
In the framework of the retarded cumulant, the effective mass at $k=0$ is
defined as:
  \begin{equation}
  \frac{m^*}{m_0}=
  \left[1+\frac{1}{\hbar}
  \frac{\partial\,\text{Re}\,\Sigma_k(\omega)}{\partial \omega}
  +\frac{\partial\,\text{Re}\,\Sigma_k(\omega)}{\partial \epsilon_k}
  \right]^{-1}_{\epsilon_k=0,\omega=0},
  \end{equation}
where $\Sigma_k(\omega)$ is now the full retarded self-energy introduced
in Eq.~\eqref{eq.sigma.t}. Following the same steps as before, one finds 
for the expansion to first order in $\alpha$:
  \begin{eqnarray}
  \frac{m^*}{m_0}&=&
  1-\frac{\alpha}{6\pi}
  \Bigg(
    \frac{8\hbar^2\omega_0^2+4E_{\rm F}^2}{E_{\rm F}^2-\Omega_0^2}
    \sqrt{\frac{\hbar\omega_0}{E_{\rm F}}}-\pi \\
    &+&2\left(
      \tan^{-1}\sqrt{\frac{E_{\rm F}}{\hbar\omega_0}}
      -\tanh^{-1}\sqrt{\frac{E_{\rm F}}{\hbar\omega_0}}
    \right)
  \Bigg)
  +\mathcal{O}\left(\alpha\right)^2\nonumber
  \end{eqnarray}
Taking the limit $E_{\rm F}\rightarrow 0$, we reach
  \begin{eqnarray}
  \lim_{\rm E_{\rm F}\rightarrow 0}\frac{m^*}{m_0}=
  \frac{4\alpha}{3\pi}\sqrt{\frac{\hbar\omega_0}{E_{\rm F}}}
  +\left(1+\frac{\alpha}{6}\right)
  +\mathcal{O}\left(E_{\rm F}\right)^{3/2},
  \end{eqnarray}
which has the same behavior for small $E_{\rm F}$ as 
Eq.~\eqref{eq.meff.cumul.ef} in the main text.

\section{Small-coupling limits, Dyson approach}\label{sec.small.coupling}
To obtain the expansion of Eq.~\eqref{eq.en.dyson.dop} to linear order in
$\alpha$, it suffices to set $E_0=0$ on the left-hand side of the equation.
We find
\nk{
  \begin{eqnarray}\label{eq.en.dyson.dop.expand}
  \frac{E_0}{\hbar\omega_0}&=&\frac{\alpha}{\pi}
  \text{Re}\Bigg[
    \log\frac{\sqrt{\Omega_0^*}+\sqrt{E_{\rm F}}}
    {\sqrt{\Omega_0^*}-\sqrt{E_{\rm F}}}
    +i\log\frac{i\sqrt{\Omega_0}+\sqrt{E_{\rm F}}}
    {i\sqrt{\Omega_0}-\sqrt{E_{\rm F}}}-\pi
    \nonumber \\
    &+&\frac{1}{2}\sqrt{\frac{\hbar\omega_0}{E_{\rm F}}}
    \Bigg(
      \text{Li}_2\frac{2\,\sqrt{E_{\rm F}}}
      {\sqrt{E_{\rm F}}+\sqrt{E_{\rm F}+\Omega_0}^*} \nonumber \\
      &+&\text{Li}_2\frac{2\,\sqrt{E_{\rm F}}}
      {\sqrt{E_{\rm F}}-\sqrt{E_{\rm F}+\Omega_0}^*}
      -\text{Li}_2\frac{2\,\sqrt{E_{\rm F}}}
      {\sqrt{E_{\rm F}}+\sqrt{E_{\rm F}-\Omega_0}} \nonumber \\
      &+&\text{Li}_2\frac{2\,\sqrt{E_{\rm F}}}
      {\sqrt{E_{\rm F}}-\sqrt{E_{\rm F}-\Omega_0}}
      +i\pi\log\frac{\sqrt{E_{\rm F}-\Omega_0}+\sqrt{E_{\rm F}}}
      {\sqrt{E_{\rm F}-\Omega_0}-\sqrt{E_{\rm F}}}
    \Bigg)
  \Bigg]\nonumber \\ &+& \mathcal{O}(\alpha)^2.
  \end{eqnarray}
}
The Dyson effective mass for small $\alpha$ is given by
\nk{
  \begin{eqnarray}\label{eq.meff.dyson.dop.expand}
  &&\frac{m^*}{m_0}=
    1+\frac{\alpha(\hbar\omega_0)^{3/2}}
    {2\pi\,\Omega_0^{3/2}}\\
    &&\times\Bigg(
      \log\frac{\sqrt{\Omega_0^*}+\sqrt{E_{\rm F}}}
      {\sqrt{\Omega_0^*}-\sqrt{E_{\rm F}}}
      +\frac{2\sqrt{E_{\rm F}\,\Omega_0^*}}{\W0^*-E_{\rm F}} \nonumber\\
    &&-i\log\frac{i\sqrt{\Omega_0}+\sqrt{E_{\rm F}}}
      {\sqrt{i\Omega_0}-\sqrt{E_{\rm F}}}
      -\frac{2\sqrt{E_{\rm F}\W0}}
      {\Omega_0^*+\ef}+\pi
    \Bigg)
  \nonumber\\
  &&
    -\frac{2\alpha}{3\pi}
    \Bigg(
      \tanh^{-1}\frac{\sqrt{E_{\rm F}}}{\sqrt{\Omega_0}}
      -\tanh^{-1}\frac{\sqrt{E_{\rm F}}}{\sqrt{-\Omega_0^*}}
      -\frac{\pi}{2}
      \nonumber\\
      &&-\frac{\sqrt{E_{\rm F}\,\hbar\omega_0}
      \left(E_{\rm F}-2\Omega_0^*\right)}
      {\left(E_{\rm F}-\Omega_0^*\right)^2}
      -\frac{\sqrt{E_{\rm F}\,\hbar\omega_0}
      \left(E_{\rm F}-2\Omega_0\right)}
      {\left(E_{\rm F}+\Omega_0\right)^2}
      \nonumber\\
      &&+\frac{\left(\w0\right)^{5/2}}{\sqrt{\ef}\left(\ef-\W0^*\right)^2}
      +\frac{\left(\w0\right)^{5/2}}{\sqrt{\ef}\left(\ef+\W0\right)^2}
    \Bigg)+\mathcal{O}(\alpha)^2\nonumber
  \end{eqnarray}
}
When we take the limit of this expression for small $E_{\rm F}$, we recover
Eq.~\eqref{eq.meff.dyson.ef}.

\section{Renormalized quantities including free-carrier screening}\label{sec.renorm.screened}
Upon including free-carrier screening, the Dyson QP energy can be 
expressed in terms of the one-dimensional integral
\nk{
  \begin{eqnarray}\label{eq.ek.dyson.scr}
  E_k&=&\epsilon_k
  -\frac{\alpha\,(\hbar\omega_0)^{3/2}}{2\pi\,\sqrt{\epsilon_k}} 
  \nonumber \\
  &\times&\text{Re}\Bigg[
    \int_0^{k_{\rm F}-k}\frac{dq}{|\varepsilon^\text{RPA}(q)|^2\,q}
    \log\frac{E_k-\frac{\hbar^2(k+q)^2}{2m_0}+\Omega_0^*}
    {E_k-\frac{\hbar^2(k-q)^2}{2m_0}+\Omega_0^*} \nonumber \\
    &+& \int_{k_{\rm F}-k}^{k_{\rm F}+k}\frac{dq}
    {|\varepsilon^\text{RPA}(q)|^2\,q}
    \log\frac{E_k-E_{\rm F}+\Omega_0^*}
    {E_k-\frac{\hbar^2(k-q)^2}{2m_0}+\Omega_0^*}
    \nonumber \\
    &+&\int_{k_{\rm F}-k}^{k_{\rm F}+k}
    \frac{dq}{|\varepsilon^\text{RPA}(q)|^2\,q}
    \log\frac{E_k-\frac{\hbar^2(k+q)^2}{2m_0}-\Omega_0}
    {E_k-E_{\rm F}-\Omega_0} \nonumber \\
    &+&\int_{k_{\rm F}+k}^\infty\frac{dq}{|\varepsilon^\text{RPA}(q)|^2\,q}
    \log\frac{E_k-\frac{\hbar^2(k+q)^2}{2m_0}-\Omega_0}
    {E_k-\frac{\hbar^2(k-q)^2}{2m_0}-\Omega_0}
  \Bigg] \nonumber \\
  &-&\text{Re}\Sigma^<_{k_{\rm F}}(E_{\rm F})
  -\text{Re}\Sigma^>_{k_{\rm F}}(E_{\rm F}).
  \end{eqnarray}
}
In particular, the occupied state at the band bottom becomes
\nk{
  \begin{eqnarray}\label{eq.en.dyson.scr}
  &&\frac{E_0}{\hbar\omega_0}=\frac{\alpha}{\pi}\,
  \text{Re}\Bigg[
    \int_0^{k_{\rm F}}\frac{dq/k_{\rm F}}
    {|\varepsilon^\text{RPA}(q)|^2}
    \frac{2\sqrt{E_{\rm F}\,\hbar\omega_0}}
    {E_0-\frac{\hbar^2q^2}{2m_0}+\Omega_0^*} \\
    &&+\int_{k_{\rm F}}^\infty
    \frac{dq/k_{\rm F}}{|\varepsilon^\text{RPA}(q)|^2}
    \frac{2\sqrt{E_{\rm F}\,\hbar\omega_0}}
    {E_0-\frac{\hbar^2q^2}{2m_0}-\Omega_0} \nonumber\\
    &&+\frac{1}{2\pi}\sqrt{\frac{\hbar\omega_0}{E_{\rm F}}}
    \int_0^{2k_{\rm F}}\frac{dq}{|\varepsilon^\text{RPA}(q)|^2\,q}\times
    \nonumber \\
    &&\hspace{-1mm}\Bigg(\hspace{-1mm}
      \log\frac{\Omega_0^*}
      {E_{\rm F}-\frac{\hbar^2(k_{\rm F}-q)^2}{2m_0}+\Omega_0^*}
      +\log\frac{E_{\rm F}-\frac{\hbar^2(k_{\rm F}+q)^2}{2m_0}-\Omega_0}
      {-\Omega_0}\hspace{-1mm}
    \Bigg) \nonumber \\
    &&+\frac{1}{2\pi}\sqrt{\frac{\hbar\omega_0}{E_{\rm F}}}
    \int_{2k_{\rm F}}^\infty\frac{dq}{|\varepsilon^\text{RPA}(q)|^2\,q}
    \log\frac{E_{\rm F}-\frac{\hbar^2(k_{\rm F}+q)^2}{2m_0}-\Omega_0}
    {E_{\rm F}-\frac{\hbar^2(k_{\rm F}-q)^2}{2m_0}-\Omega_0}
  \Bigg].\nonumber 
  \end{eqnarray}
}
The effective mass at the zone center is given by:
\nk{
  \begin{eqnarray}\label{eq.meff.dyson.scr}
  &&\frac{m^*}{m_0}=
  \text{Re}\Bigg[
    1+\frac{2\,\alpha\,\left(\w0\right)^{3/2}}{\pi\,\sqrt\ef}\nonumber \\
    &&\times\Bigg(
      \int_0^{k_{\rm F}}\frac{dq/\kf}{|\varepsilon^\text{RPA}(q)|^2}
      \frac{\ef}
      {\left(E_0-\frac{\hbar^2q^2}{2m_0}+\Omega_0^*\right)^2} \nonumber \\
      &&+\int_{k_{\rm F}}^\infty\frac{dq/\kf}{|\varepsilon^\text{RPA}(q)|^2}
    \frac{\ef}{\left(E_0-\frac{\hbar^2q^2}{2m_0}-\Omega_0\right)^2} 
  \Bigg] \nonumber \\
  &&\times\Bigg[
    1+\frac{2\alpha\,(\hbar\omega_0)^{3/2}}
    {3\pi\,\sqrt{E_{\rm F}}}\nonumber \\
    &&\times\Bigg(
      \int_0^{k_{\rm F}}\frac{dq/\kf}{|\varepsilon^\text{RPA}(q)|^2}
      \frac{\ef\,\left(3E_0+\frac{\hbar^2 q^2}{2m_0}+3\Omega_0^*\right)}
      {\left(E_0-\frac{\hbar^2q^2}{2m_0}+\Omega_0^*\right)^3}\nonumber \\
      &&+\int_{k_{\rm F}}^\infty\frac{dq/\kf}{|\varepsilon^\text{RPA}(q)|^2}
      \frac{\ef\,\left(3E_0+\frac{\hbar^2 q^2}{2m_0}-3\Omega_0\right)}
      {\left(E_0-\frac{\hbar^2q^2}{2m_0}-\Omega_0\right)^3}
      \nonumber\\
      &&-\frac{1}{|\varepsilon^\text{RPA}(\kf)|^2}
      \left(
        \frac{\W0^*+E_0}{\left(E_0-\ef+\W0^*\right)^2}
        +\frac{\W0-E_0}{\left(E_0-\ef-\W0\right)^2}
      \right)
    \Bigg)
  \Bigg]^{-1}.\nonumber\\
  \end{eqnarray}
}
To linear order in $\alpha$, the weak-coupling limit of 
Eq.~\eqref{eq.en.dyson.scr} is simply given by:
\nk{
  \begin{eqnarray}\label{eq.en.dyson.scr.expand}
  &&\frac{E_0}{\hbar\omega_0}=\frac{\alpha}{\pi}
  \text{Re}\Bigg[
    \int_0^{k_{\rm F}}\frac{dq/k_{\rm F}}{|\varepsilon^\text{RPA}(q)|^2}
    \frac{2\sqrt{E_{\rm F}\,\hbar\omega_0}}
    {-\frac{\hbar^2q^2}{2m_0}+\Omega_0^*} \nonumber \\
    &&+\int_{k_{\rm F}}^\infty
    \frac{dq/k_{\rm F}}{|\varepsilon^\text{RPA}(q)|^2}
    \frac{2\sqrt{E_{\rm F}\,\hbar\omega_0}}
    {-\frac{\hbar^2q^2}{2m_0}-\Omega_0} \nonumber \\
    &&+\frac{1}{2}\sqrt{\frac{\hbar\omega_0}{E_{\rm F}}}
    \int_0^{2k_{\rm F}}\frac{dq}{|\varepsilon^\text{RPA}(q)|^2\,q}\times 
    \nonumber \\
    &&\hspace{-1mm}\Bigg(\hspace{-1mm}
      \log\frac{\Omega_0^*}
      {E_{\rm F}-\frac{\hbar^2(k_{\rm F}-q)^2}{2m_0}+\Omega_0^*}
      +\log\frac{E_{\rm F}-\frac{\hbar^2(k_{\rm F}+q)^2}{2m_0}-\Omega_0}
      {-\Omega_0}\hspace{-1mm}
    \Bigg) \nonumber \\
    &&+\frac{1}{2}\sqrt{\frac{\hbar\omega_0}{E_{\rm F}}}
    \int_{2k_{\rm F}}^\infty\frac{dq}{|\varepsilon^\text{RPA}(q)|^2\,q}
    \log\frac{E_{\rm F}-\frac{\hbar^2(k_{\rm F}+q)^2}{2m_0}-\Omega_0}
    {E_{\rm F}+\frac{\hbar^2(k_{\rm F}-q)^2}{2m_0}-\Omega_0}
  \Bigg] \nonumber \\
  &&+\mathcal{O}(\alpha^2).
  \end{eqnarray}
}
For Eq.~\eqref{eq.meff.dyson.scr}, we find at small $\alpha$:
\nk{
  \begin{eqnarray}\label{eq.meff.dyson.scr.expand}
  \frac{m^*}{m_0}&=&
    1+\frac{2\,\alpha\,\left(\w0\right)^{3/2}}{3\pi\,\sqrt\ef}
    \text{Re}\Bigg[
      \int_0^{k_{\rm F}}\frac{dq/\kf}{|\varepsilon^\text{RPA}(q)|^2}
      \frac{3\ef}{\left(\frac{\hbar^2q^2}{2m_0}-\W0^*\right)^2} \nonumber \\
      &+&\int_{k_{\rm F}}^\infty\frac{dq/\kf}{|\varepsilon^\text{RPA}(q)|^2}
      \frac{3\ef}{\left(\frac{\hbar^2q^2}{2m_0}+\W0\right)^2} \nonumber \\
      &-&\int_0^{k_{\rm F}}\frac{dq/\kf}{|\varepsilon^\text{RPA}(q)|^2}
      \frac{\ef\,\left(\frac{\hbar^2q^2}{2m_0}+3\Omega_0^*\right)}
      {\left(\frac{\hbar^2q^2}{2m_0}-\Omega_0^*\right)^3} \nonumber \\
      &-&\int_{k_{\rm F}}^\infty\frac{dq/\kf}{|\varepsilon^\text{RPA}(q)|^2}
      \frac{\ef\left(\frac{\hbar^2q^2}{2m_0}-3\Omega_0\right)}
      {\left(\frac{\hbar^2q^2}{2m_0}+\Omega_0\right)^3}\nonumber\\
      &-&\frac{1}{|\varepsilon^\text{RPA}(\kf)|^2}
      \left(
      \frac{\W0^*}{\left(\ef-\W0^*\right)^2}
      +\frac{\W0}{\left(\ef+\W0\right)^2}
      \right)
  \Bigg]\nonumber \\
  &+&\mathcal{O}(\alpha^2).
  \end{eqnarray}
}
In the case of the cumulant approach, the QP energy for $k<k_{\rm F}$ is 
given by
  \begin{eqnarray}\label{eq.ek.cumul.scr}
  E_k&=&\epsilon_k
  -\frac{\alpha\,(\hbar\omega_0)^{3/2}}{2\pi\,\sqrt{\epsilon_k}} 
  \nonumber \\
  &\times&\text{Re}\Bigg[
	  \int_0^{k_{\rm F}-k}\frac{dq}{|\varepsilon^\text{RPA}(q)|^2\,q}
    \log\frac{\Omega_0-\frac{\hbar^2(q^2+2kq)}{2m_0}}
    {\Omega_0-\frac{\hbar^2(q^2-2kq)}{2m_0}} \nonumber \\
	  &+& \int_{k_{\rm F}-k}^{k_{\rm F}+k}\frac{dq}{|\varepsilon^\text{RPA}(q)|^2\,q}
	  \log\frac{\Omega_0+\epsilon_k-E_{\rm F}}
    {\Omega_0-\frac{\hbar^2(q^2-2kq)}{2m_0}}
  \Bigg] \nonumber \\
  &-&\text{Re}\Sigma^<_{k_{\rm F}}(E_{\rm F}).
  \end{eqnarray}
In the limit $k\rightarrow 0$, we recover
  \begin{eqnarray}\label{eq.en.cumul.scr}
  &&\frac{E_0}{\hbar\omega_0}=\frac{\alpha}{\pi}
  \text{Re}\Bigg[
    \int_0^{k_{\rm F}}
    \frac{dq/k_{\rm F}}{|\varepsilon^\text{RPA}(q)|^2}
    \frac{2\sqrt{E_{\rm F}\,\hbar\omega_0}}
    {\Omega_0-\frac{\hbar^2q^2}{2m_0}} \\
    &&+\frac{1}{2}\sqrt{\frac{\hbar\omega_0}{E_{\rm F}}}
    \int_0^{2k_{\rm F}}\frac{dq}{|\varepsilon^\text{RPA}(q)|^2\,q}
    \log\frac{\Omega_0}{E_F-\frac{\hbar^2(k_{\rm F}-q)^2}{2m_0}+\Omega_0}
  \Bigg], \nonumber
  \end{eqnarray}
which is already linear in $\alpha$. Lastly, the cumulant effective mass 
at $\Gamma$ is given by:
  \begin{eqnarray}\label{eq.meff.cumul.scr}
  &&\frac{m^*}{m_0}=
  \text{Re}\Bigg[
    1-\frac{2\alpha\,(\hbar\omega_0)^{3/2}}
    {\pi\,k_{\rm F}/\sqrt{E_{\rm F}}}\nonumber\\
    &&\times\int_0^{k_{\rm F}}\frac{dq}{|\varepsilon^\text{RPA}(q)|^2}
    \Bigg[\hspace{-0.3mm}
      \frac{1}{\left(-\frac{\hbar^2q^2}{2m_0}+\Omega_0\right)^2}
      +\frac{\frac{\hbar^2q^2}{2m_0}+3\Omega_0}
      {3\left(\frac{\hbar^2q^2}{2m_0}-\Omega_0\right)^3}\hspace{-0.5mm}
    \Bigg] \nonumber \\
    &&-\frac{2\alpha\,\hbar\omega_0}{3\pi}
    \sqrt{\frac{\hbar\omega_0}{E_{\rm F}}}
    \frac{1}{|\varepsilon^\text{RPA}(k_{\rm F})|^2}
    \frac{\Omega_0}{\left(-E_{\rm F}+\Omega_0\right)^2}
    \Bigg]^{-1},
  \end{eqnarray}
and its small-$\alpha$ expansion is equal to
  \begin{eqnarray}\label{eq.meff.cumul.scr.expand}
  &&\frac{m^*}{m_0}=
  \text{Re}\Bigg[
    1+\frac{2\alpha\,(\hbar\omega_0)^{3/2}}
    {\pi\,k_{\rm F}/\sqrt{E_{\rm F}}}\\
    &&\times\int_0^{k_{\rm F}}\frac{dq}{|\varepsilon^\text{RPA}(q)|^2}
    \Bigg[\hspace{-0.3mm}
      \frac{1}{\left(-\frac{\hbar^2q^2}{2m_0}+\Omega_0\right)^2}
      +\frac{\frac{\hbar^2q^2}{2m_0}+3\Omega_0}
      {3\left(\frac{\hbar^2q^2}{2m_0}-\Omega_0\right)^3}\hspace{-0.5mm}
    \Bigg] \nonumber \\
    &&+\frac{2\alpha\,\hbar\omega_0}{3\pi}
    \sqrt{\frac{\hbar\omega_0}{E_{\rm F}}}
    \frac{1}{|\varepsilon^\text{RPA}(k_{\rm F})|^2}
    \frac{\Omega_0}{\left(-E_{\rm F}+\Omega_0\right)^2}
  \Bigg]+\mathcal{O}\left(\alpha^2\right).\nonumber
  \end{eqnarray}

\clearpage
\newpage
\bibliography{bibliography}{}
\clearpage

  \begin{table*}
  \centering
  \caption{Overview of equations for all quantities derived
  in this work: 
  $k$-dependent QP
  energy $E_k$, QP energy at the zone center $E_0/\hbar\omega_0$, and 
  renormalized effective mass $m^*/m_0$ for the cases of a single electron
  (\F\, polaron problem), finite doping, and finite doping
  including free-carrier screening.}
  \renewcommand{\arraystretch}{1.3}
  \begin{tabular}{p{0.2\textwidth}|P{0.08\textwidth}P{0.08\textwidth}
  P{0.08\textwidth}|P{0.08\textwidth}P{0.08\textwidth}|P{0.08\textwidth}
  P{0.08\textwidth}P{0.08\textwidth}}
  \hline 
  \hline 
  &
  \multicolumn{3}{P{0.26\textwidth}|}{Dyson} & 
  \multicolumn{2}{P{0.17\textwidth}|}{Dyson $\mathcal{O}(\alpha)$} &  
  \multicolumn{3}{P{0.26\textwidth} }{Cumulant} \\[2pt]
  &$E_k$ & $E_0/\hbar\omega_0$ & $m^*/m_0$ & 
  $E_0/\hbar\omega_0$ & $m^*/m_0$ &
  $E_k$ & $E_0/\hbar\omega_0$ & $m^*/m_0$ \\[2pt]
  \hline 
  Single electron & \eqref{eq.ek.undoped} & \eqref{eq.en.dyson.undop} &
  \eqref{eq.mass.dyson.undop} & \eqref{eq.en.dyson.undop.expand} &
  \eqref{eq.mass.dyson.undop.expand} & \eqref{eq.ek.cumul.undop} & 
  \eqref{eq.en.cumul.undop} & \eqref{eq.mass.cumul.undop} \\
  Finite doping& \eqref{eq.ek.dyson.dop} & \eqref{eq.en.dyson.dop} &
  \eqref{eq.meff.dyson.dop} & \eqref{eq.en.dyson.dop.expand} &
  \eqref{eq.meff.dyson.dop.expand} & \eqref{eq.ek.cumul.dop} &
  \eqref{eq.en.cumul.dop} & \eqref{eq.meff.cumul.dop} \\
  Doping and screening & \eqref{eq.ek.dyson.scr} & 
  \eqref{eq.en.dyson.scr} & \eqref{eq.meff.dyson.scr} & 
  \eqref{eq.en.dyson.scr.expand} & \eqref{eq.meff.dyson.scr.expand} &
  \eqref{eq.ek.cumul.scr} & \eqref{eq.en.cumul.scr} & 
  \eqref{eq.meff.cumul.scr} \\[2pt]
  \hline
  \hline
  \end{tabular}
  \renewcommand{\arraystretch}{1}
  \label{tab1}
  \end{table*}

\begin{table}
	\centering
	\renewcommand{\arraystretch}{1.3}
	\caption{Comparison of calculated and experimental quasi-particle weights}
	\label{tab2}
	\begin{tabular}{c|c|c|c}
		\hline
		\hline
		& Experiment \cite{WangZ2016} & Dyson & Cumulant \\
		\hline
		QP weight $Z_{k=0} $& 0.2 & 0.38 & 0.31\\
		\hline
		\hline
	\end{tabular}
\end{table}

  \begin{figure*}[t]
  \begin{minipage}{0.45\linewidth}
  \centering
  \includegraphics[width=\linewidth]{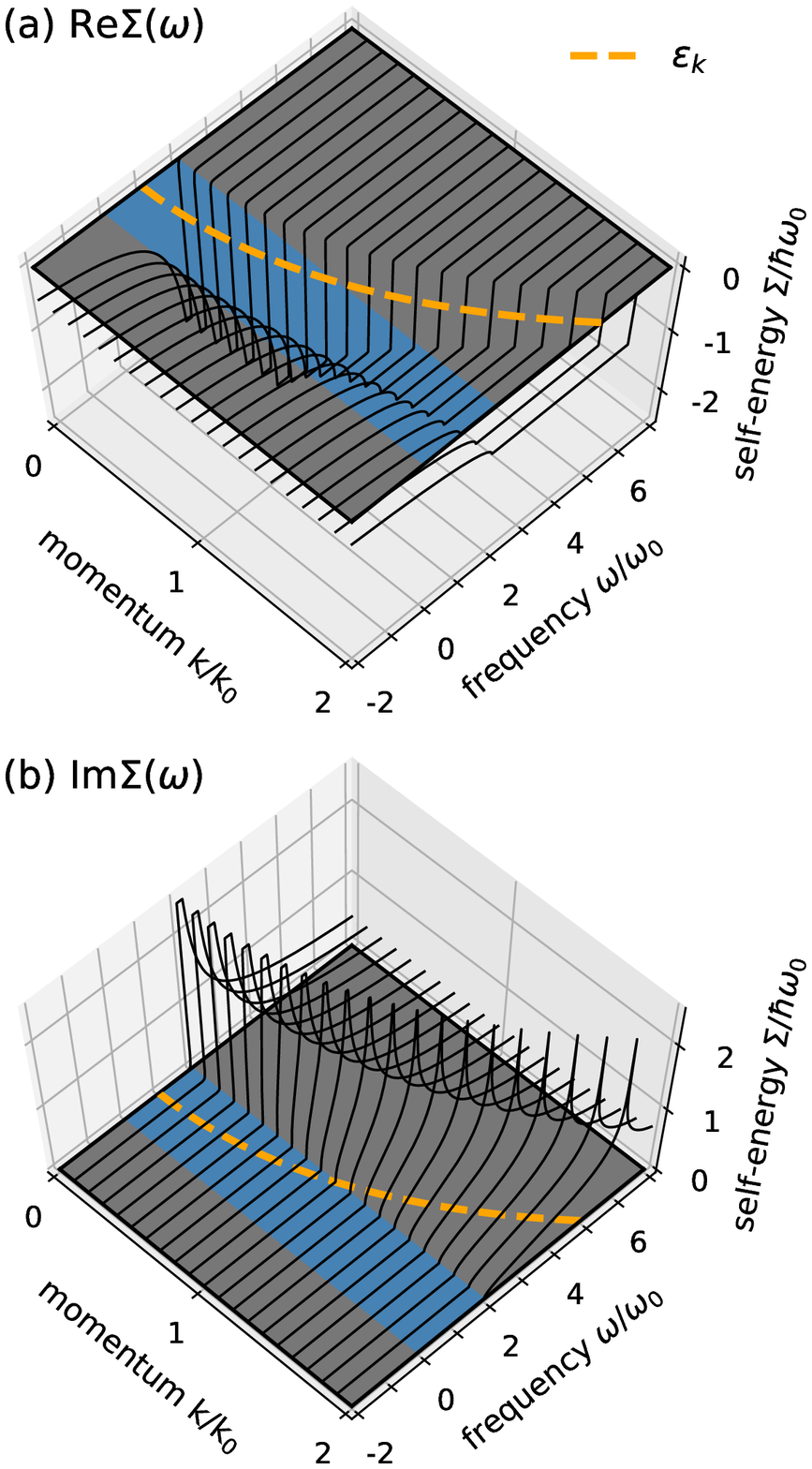}
  \end{minipage}
  \hfill
  \begin{minipage}{0.54\linewidth}
  \centering
  \includegraphics[width=\linewidth]{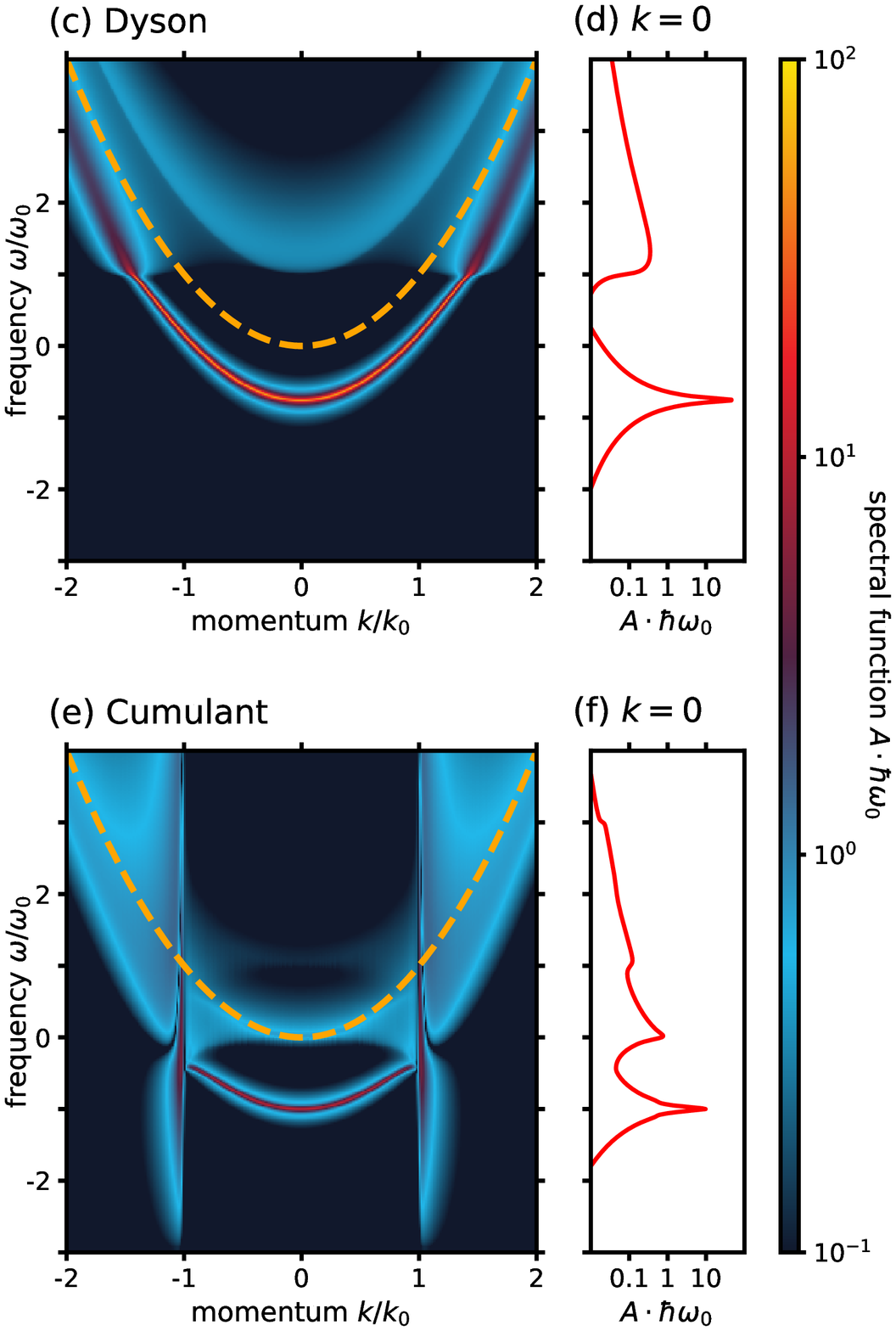}
  \end{minipage}
  \caption{
  Self-energy and spectral function for the undoped, empty-band \F\ model
  with $\alpha=1$.
  (a) Real part of the greater self-energy 
  (black lines) relative to the dispersion of the non-interacting electron
  in units of the phonon energy $\hbar\omega_0$ (dashed orange line). The
  blue area indicates the energy range 
  $\left[\epsilon_0-\hbar\omega_0,\epsilon_0+\hbar\omega_0\right]$.
  (b) Imaginary part of the greater self-energy.
  (c) Color plot of the Dyson spectral function of the undoped system. 
  The dashed orange line indicates the dispersion of the non-interacting
  electron in units of the phonon energy $\hbar\omega_0$.
  (d) Logarithmic line plot of the Dyson spectral function at the 
  $\Gamma$ point.
  (e) Color plot of the second-order cumulant spectral function.
  (f) Logarithmic line plot of the cumulant spectral function at the
  $\Gamma$ point.}
  \label{fig2}
  \end{figure*}

  \begin{figure}[t]
  \centering
  \includegraphics[width=\linewidth]{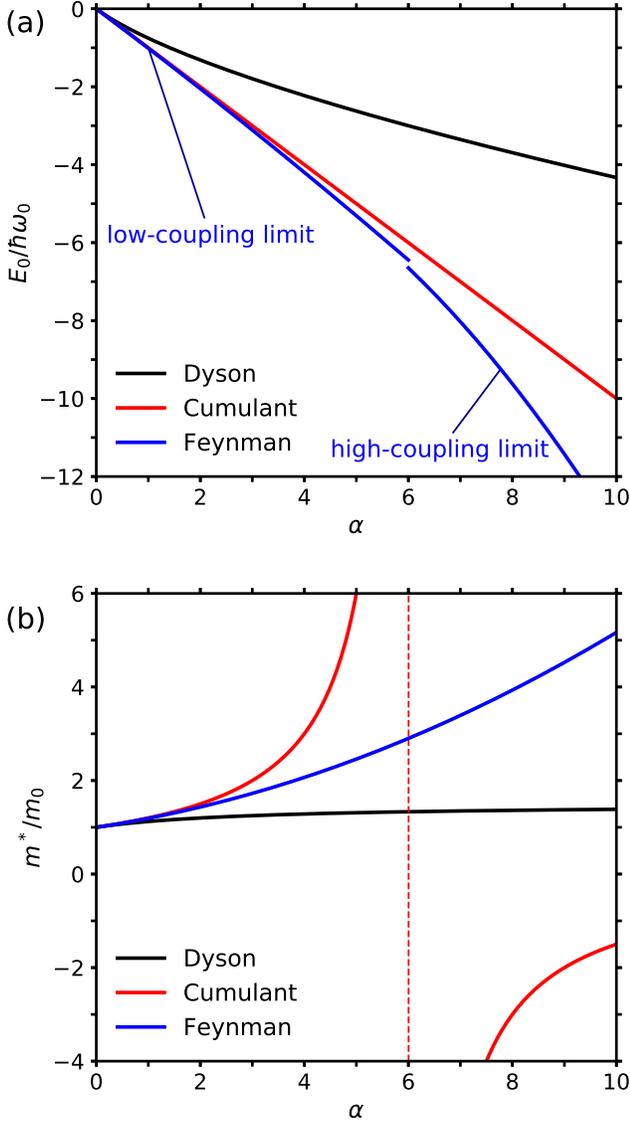}
  \caption{
  (a) Electron energy renormalization in the undoped \F\ model as a function of the 
  coupling strength $\alpha$. Shown in blue are the weak- and 
  strong-coupling limits of the Feynman model.
  (b) Electron mass renormalization in the undoped \F\,model as a function of the
  coupling strength $\alpha$.}
  \label{fig3}
  \end{figure}

  \begin{figure*}[t]
  \begin{minipage}{0.45\linewidth}
  \centering
  \includegraphics[width=\linewidth]{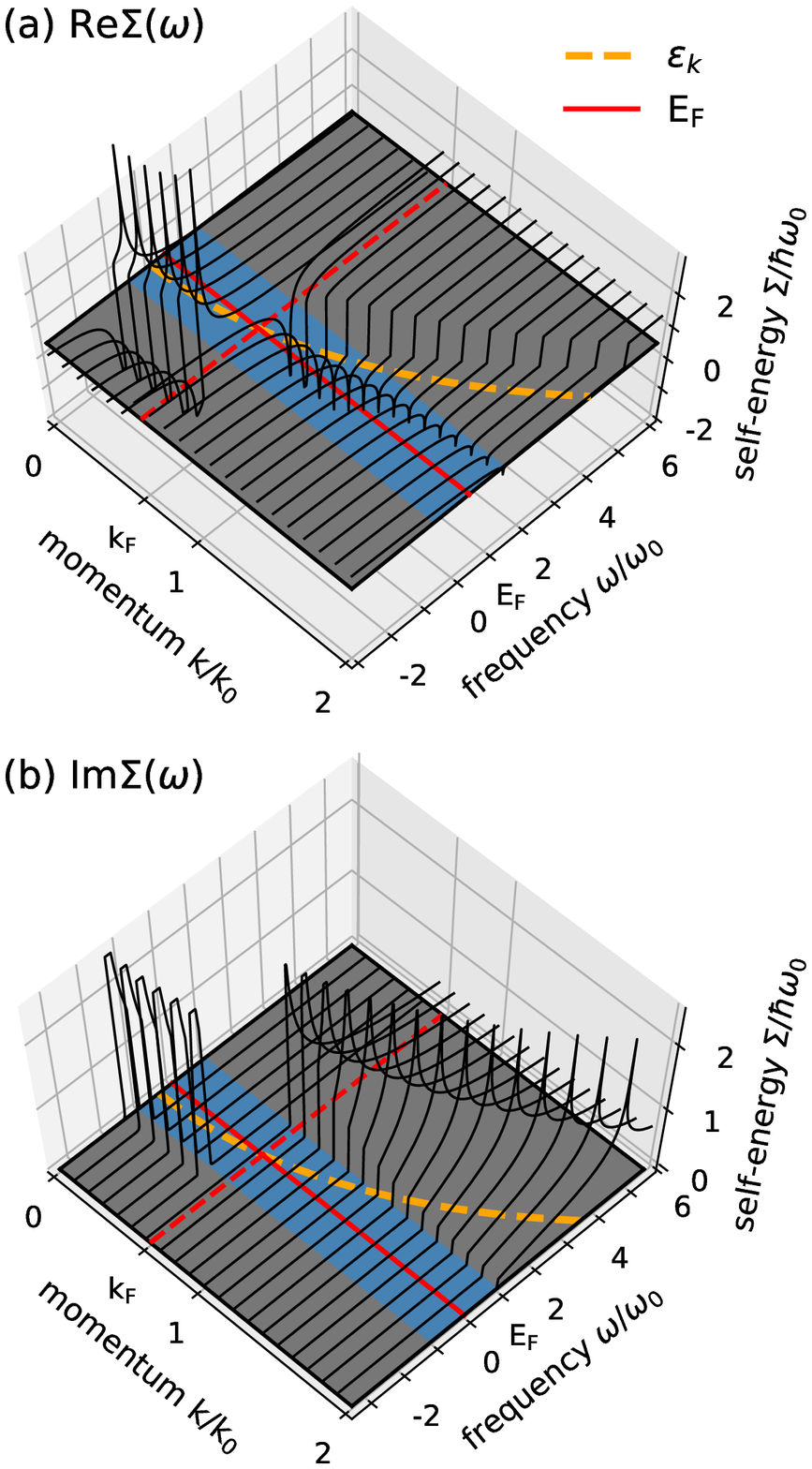}
  \end{minipage}
  \hfill
  \begin{minipage}{0.54\linewidth}
  \centering
  \includegraphics[width=\linewidth]{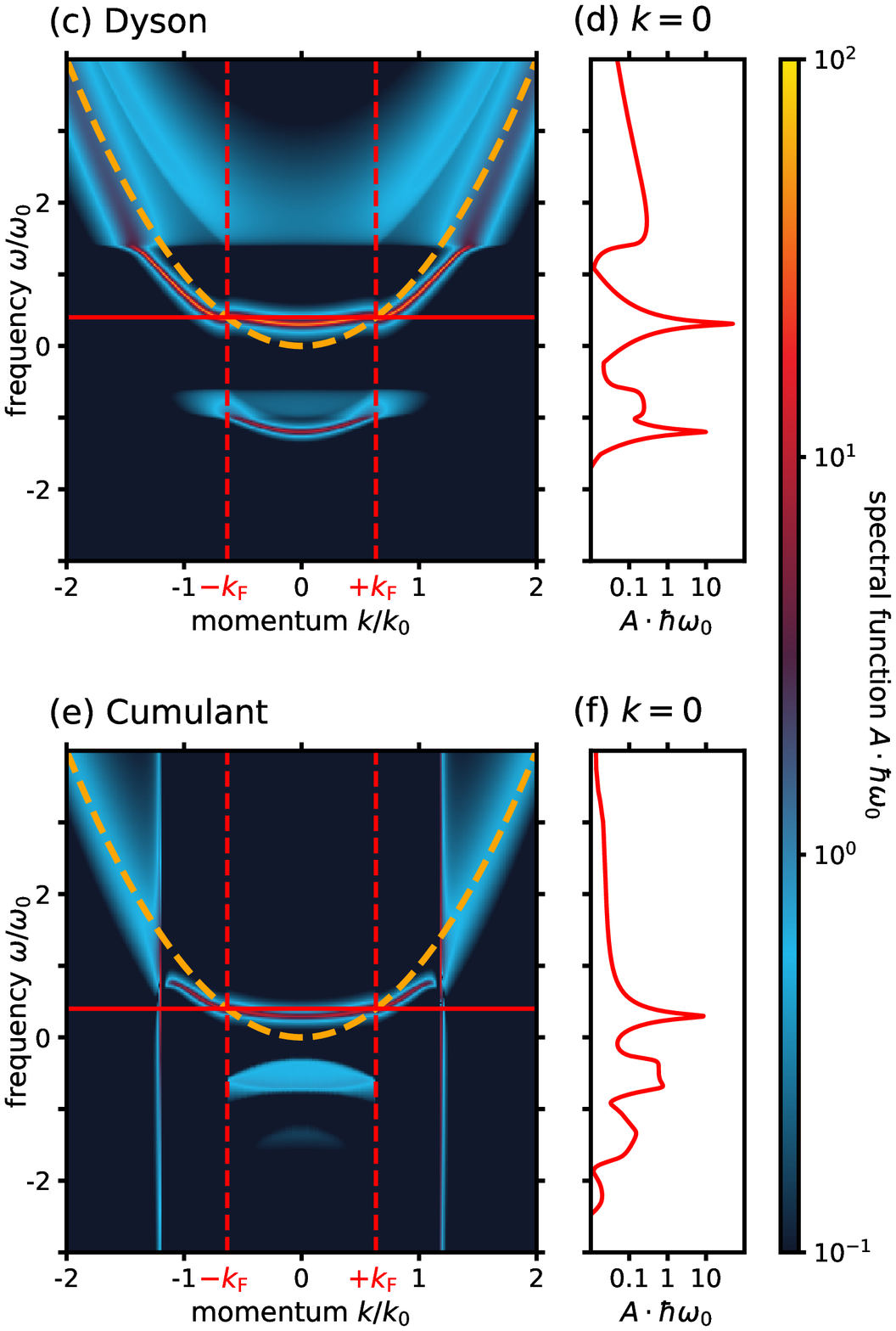}
  \end{minipage}
  \caption{
  Self-energy and spectral function for the \F\ model with $\alpha=1$ and 
  a Fermi energy of $E_{\rm F}/\hbar\omega_0=0.4$. Free-carrier screening
 is not included.
  (a) Real part of the lesser and greater self-energies in the 
  extreme anti-adiabatic limit (black lines) relative to the dispersion
  of the non-interacting particle (dashed orange line). The solid and 
  dashed red lines indicate the Fermi energy and Fermi momentum, 
  respectively; the blue area indicates the energy range 
  $\left[E_{\rm F}-\hbar\omega_0,E_{\rm F}+\hbar\omega_0\right]$.
  (b) Imaginary part of the self-energy.
  (c) Color plot of the Dyson spectral function in the extreme
  anti-adiabatic limit relative to the non-interacting electron energy
  (dashed orange line) and the Fermi energy (red line).
  (d) Logarithmic line plot of the Dyson spectral function at the 
  $\Gamma$ point.
  (e) Color plot of the second-order cumulant spectral function.
  (f) Logarithmic line plot of the cumulant spectrum at $k=0$.}
  \label{fig4}
  \end{figure*}
  
  \begin{figure*}
  	\centering
  	\includegraphics[width=\linewidth]{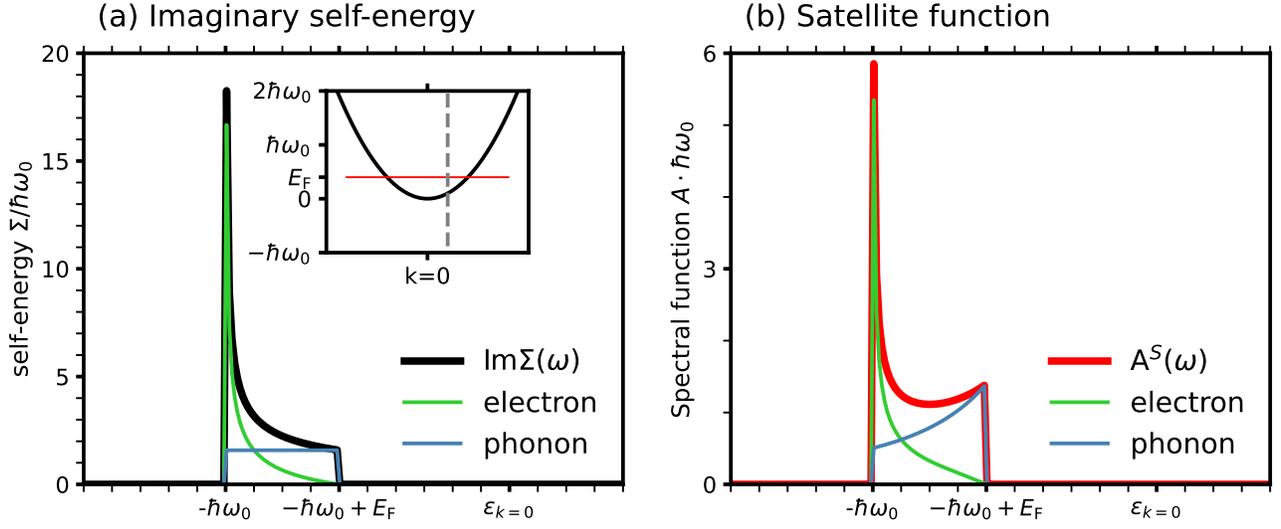}
  	\caption{(a) Imaginary lesser self-energy at $\epsilon_k=0.1\hbar\omega_0$ at a Fermi energy 
  	of $\ef=0.4\hbar\omega_0$ and $\alpha=1$ (black line). 
    Contributions to Eq.~\eqref{eq.dop.sigma.l} arising from the singularity in the non-interacting 
    \textsl{electron} Green's function are shown in green, those due to \textsl{phonon} emission and
    absorption processes across the entire Fermi sea appear in blue.
    (b) Corresponding satellite function $A^\text{S}(\omega)$ (before convolution with the QP
    function, shown in red). Contributions to the satellite function due to the electron Green's function
    and phonon emission and absorption processes are shown in green and blue, respectively.}
	\label{fig.new}
  \end{figure*}
  
  \begin{figure*}[t]
  \begin{minipage}{0.49\linewidth}
  \centering
  \includegraphics[width=\linewidth]{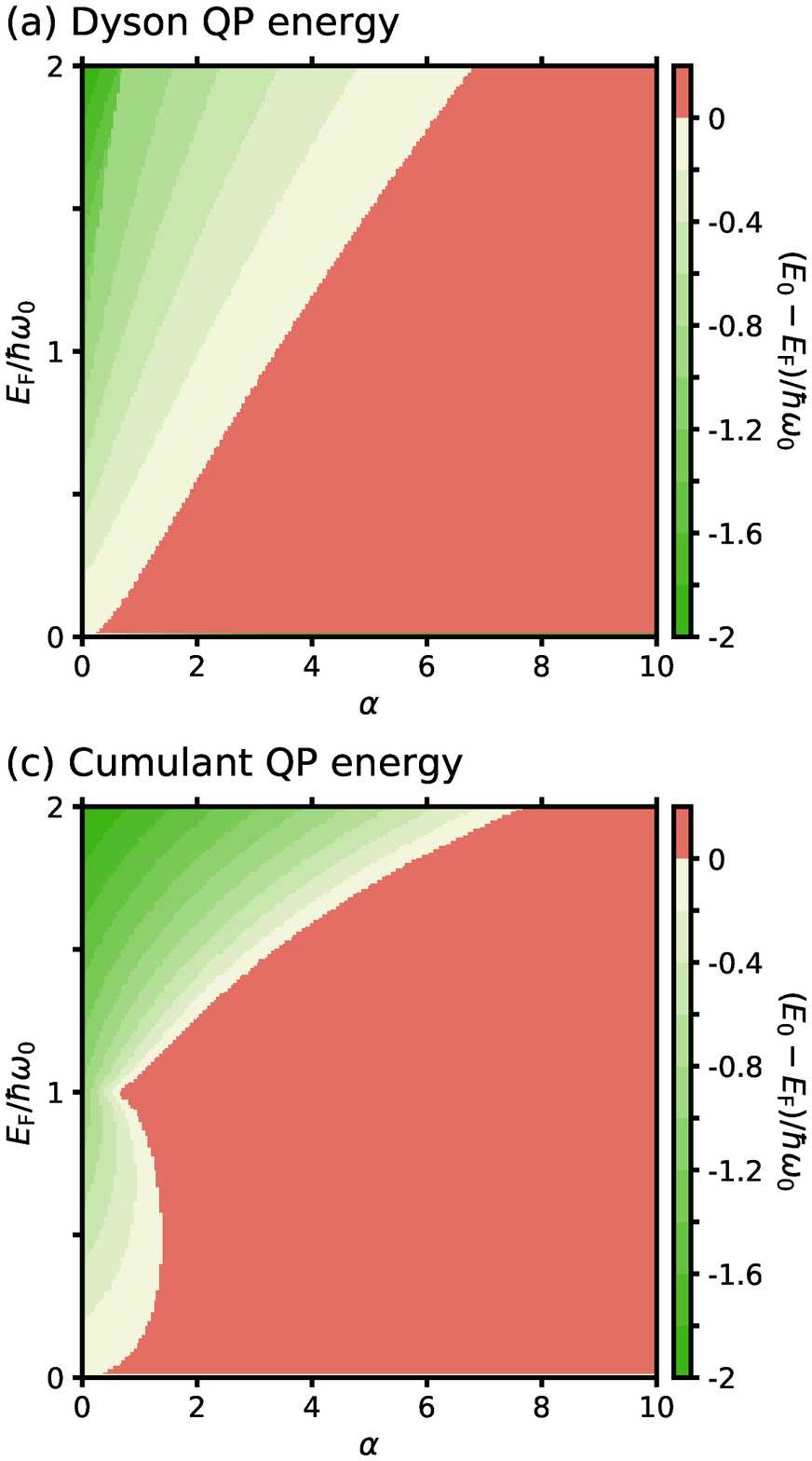}
  \end{minipage}
  \hfill
  \begin{minipage}{0.49\linewidth}
  \centering
  \includegraphics[width=\linewidth]{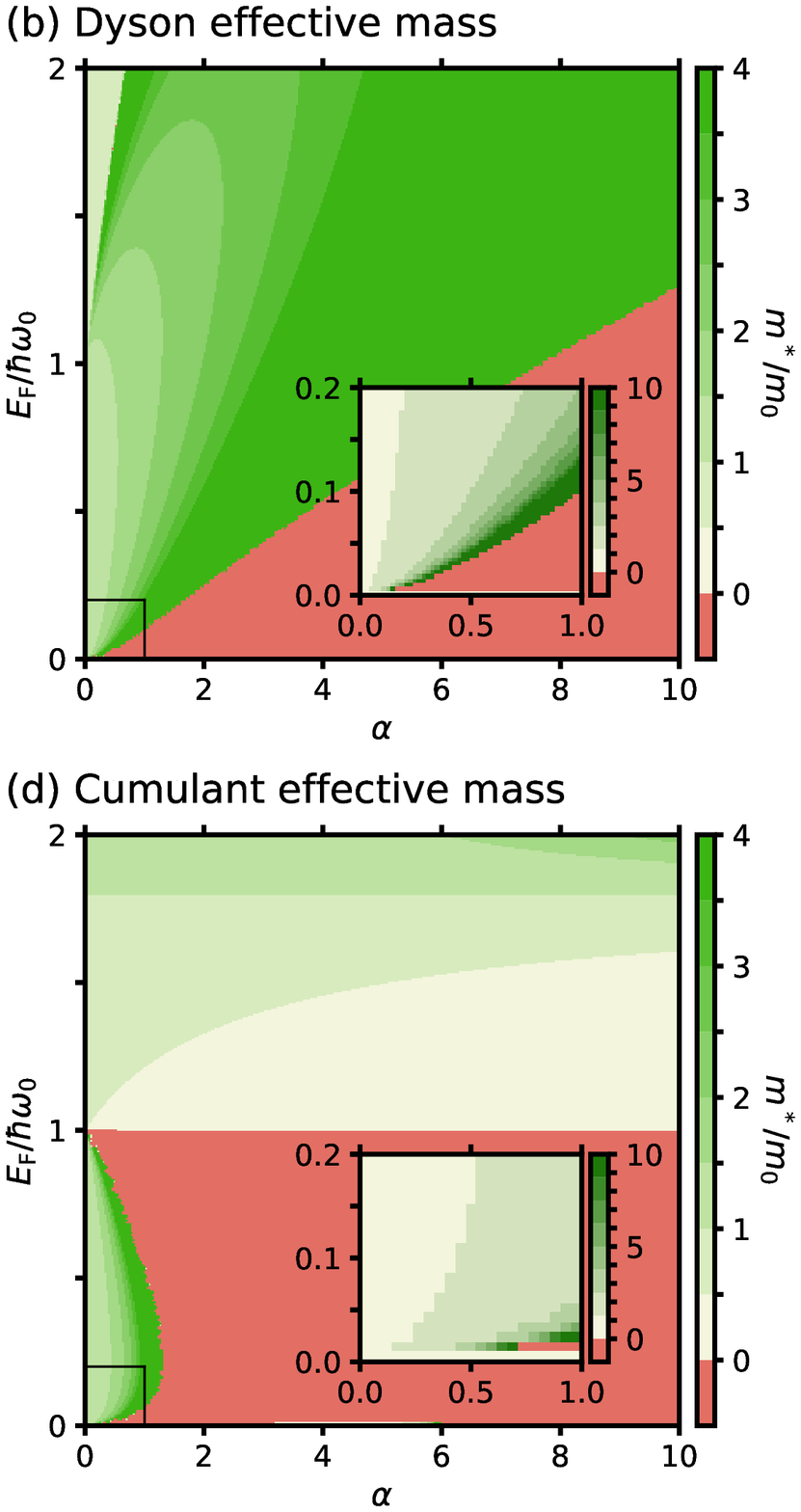}
  \end{minipage}
  \caption{
  (a) Renormalized Dyson QP energy $E_0/\hbar\omega_0$ relative to the 
  Fermi level. Negative QP energies $E_0$ (shown in green) indicate a higher
  binding energy of the interacting system, whereas positive energies
  (shown in red) imply that the renormalized QP energy lies 
  above the Fermi level, and indicate a breakdown of the Fermi surface.
  (b) Renormalized Dyson effective mass at the $\Gamma$ point. Positive values of
  $m^*$ (shown in green) indicate the renormalization of the 
  QP mass; negative values of $m^*$ (shown in red) indicate that the curvature
  of the QP spectrum at $k=0$ has become negative, and signal a breakdown
  of the second-order expansion of the self-energy. The inset shows an enlarged view for
  $\alpha<1$ and $E_{\rm F}<0.2\hbar\omega_0$; note that the color scale in the 
  inset has been extended to $m^*/m_0=10$.
  (c) Renormalized second-order cumulant QP energy $E_0/\hbar\omega_0$ 
  relative to the Fermi level.
  (d) Renormalized second-order cumulant effective mass.}
  \label{fig5}
  \end{figure*}
  
  \begin{figure}[t]
  \centering
  \includegraphics[width=\linewidth]{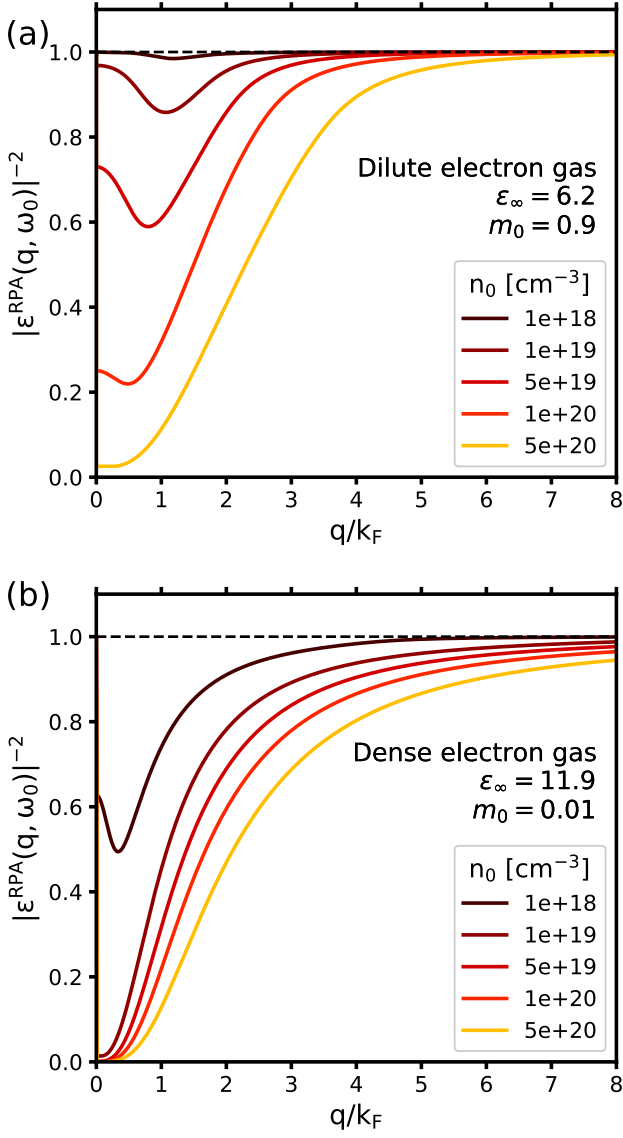}
  \caption{
  Effect of free-carrier screening on the electron-phonon coupling
  matrix element $g$ for a range of typical doping values for 
  semiconductors. 
  (a) Screening function $|\varepsilon^\text{RPA}|^{-2}$ vs. wavenumber $q$, evaluated at the phonon frequency 
  $\omega_0$ for a dilute electron gas. The parameters $\varepsilon_0$ and $m_0$ given in the
  legend correspond to
  SrTiO$_3$. The dashed line at $|\varepsilon^\text{RPA}|^{-2} = 1$
  indicates the limit of no free-carrier screening.
  (b) Same as (a), but for a dense electron gas. The
 parameters $\varepsilon_\infty$ and $m_0$ given in the legend correspond to 
  GaAs.
  }
  \label{fig6}
  \end{figure}

  \begin{figure*}[t]
  \centering
  \includegraphics[width=\linewidth]{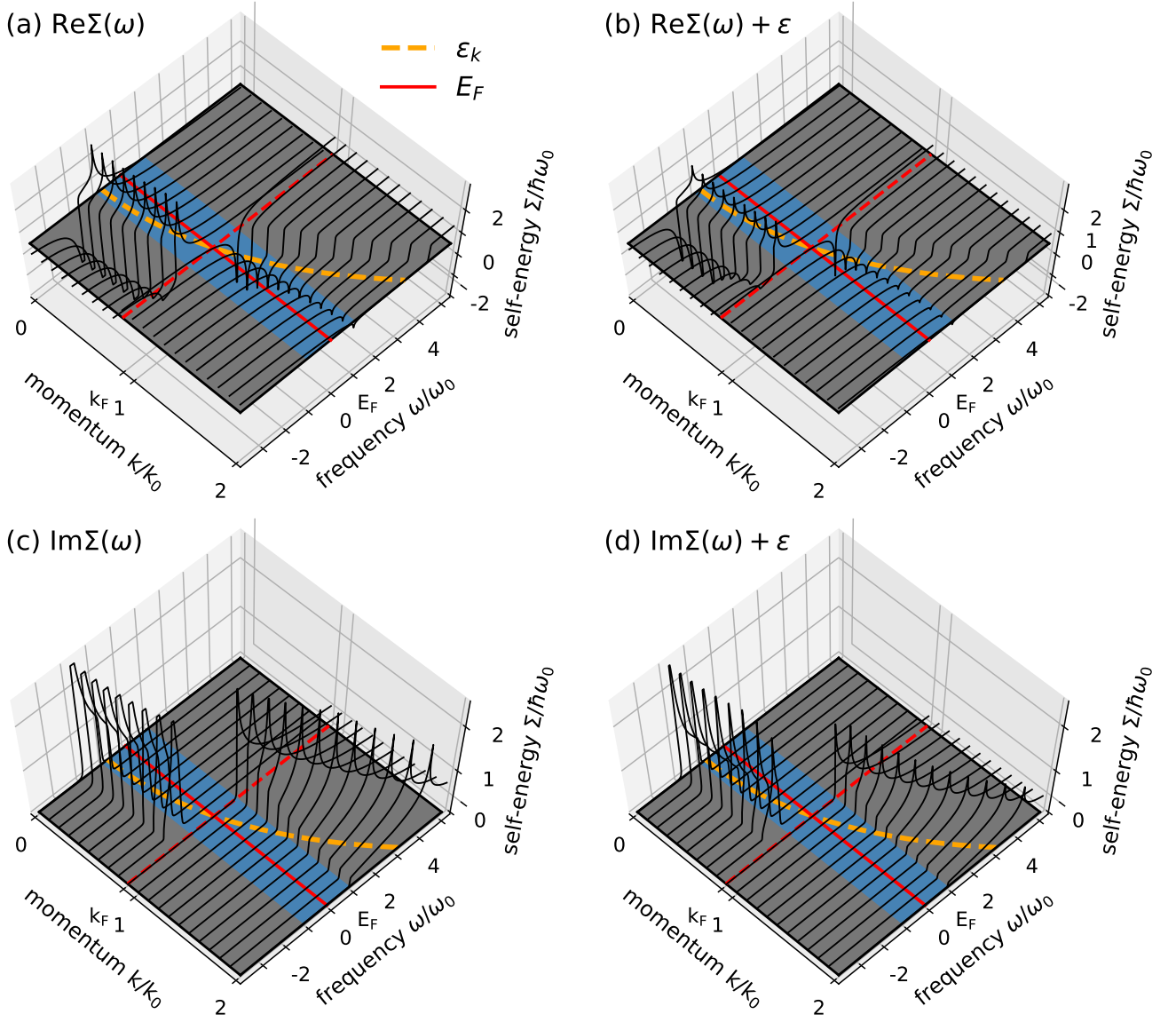}
  \caption{
  Self-energy for the \F\, model with $\alpha=1$ and a Fermi energy of $E_{\rm F}/\hbar\omega_0$=0.8 without 
  free-carrier screening $\left[\Sigma(\omega)\right]$ and with free carrier screening effects $\left[\Sigma(\omega)+\varepsilon\right]$, respectively.
  (a) Real part of the lesser and greater self-energies without free-carrier screening (black lines) relative to the 
  dispersion of the non-interacting particle (dashed orange line). The solid and dashed red lines indicate the Fermi 
  energy and Fermi momentum, respectively. The blue area indicates the energy range 
  $\left[E_{\rm F}-\hbar\omega_0,E_{\rm F}+\hbar\omega_0\right]$.
  (b) Real part of the lesser and greater self-energies including free-carrier screening, assuming a dilute electron gas
  like in the conduction band of SrTiO$_3$.
  (c) Imaginary part of the self-energy without free-carrier screening.
  (d) Imaginary part of the self-energy including free-carrier screening}
  \label{fig7}
  \end{figure*}

  \begin{figure*}[t]
  \centering
  \includegraphics[width=\linewidth]{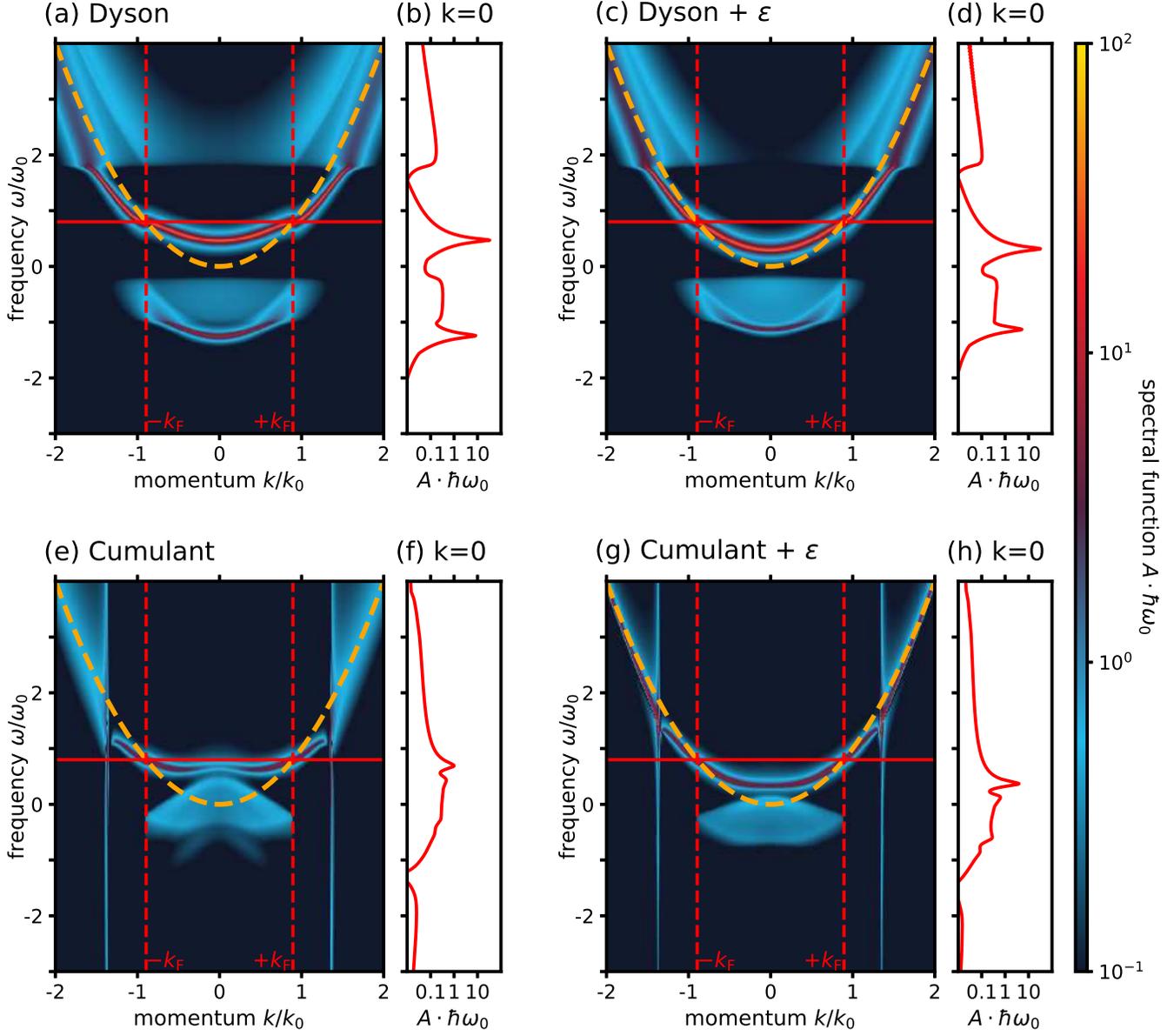}
  \caption{
  (a) Unscreened Dyson spectral function at $\alpha=1$ and 
  $E_{\rm F}/\hbar\omega_0=0.8$. The non-interacting electron energy is 
  indicated by the dashed gray line, the Fermi level by the solid red line.
  (b) Unscreened second-order cumulant spectral function. 
  (c) Logarithmic line plot of the Dyson spectral weight function at 
  $k=0$.
  (d) Logarithmic line plot of the cumulant spectral function at $k=0$.
  (e) Screened Dyson spectral function assuming a dilute electron gas like in the conduction band of SrTiO$_3$.
  (f) Screened Cumulant spectral function.
  (g) Logarithmic line plot of the screened Dyson spectral function
  (h) Logarithmic line plot of the screened cumulant spectral function.}
  \label{fig8}
  \end{figure*}

  \begin{figure*}[t]
  \begin{minipage}{0.49\linewidth}
  \centering
  \includegraphics[width=\linewidth]{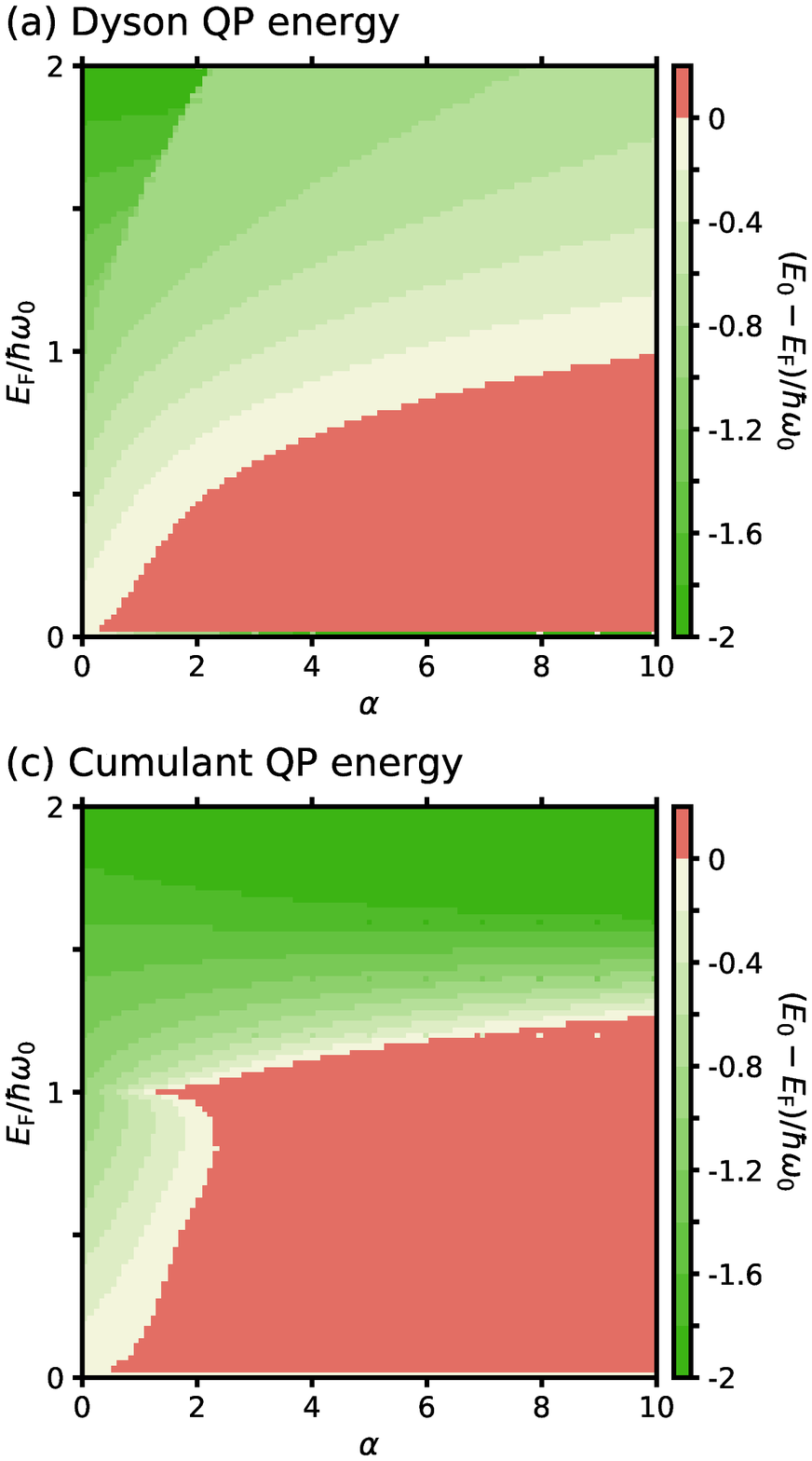}
  \end{minipage}
  \hfill
  \begin{minipage}{0.49\linewidth}
  \centering
  \includegraphics[width=\linewidth]{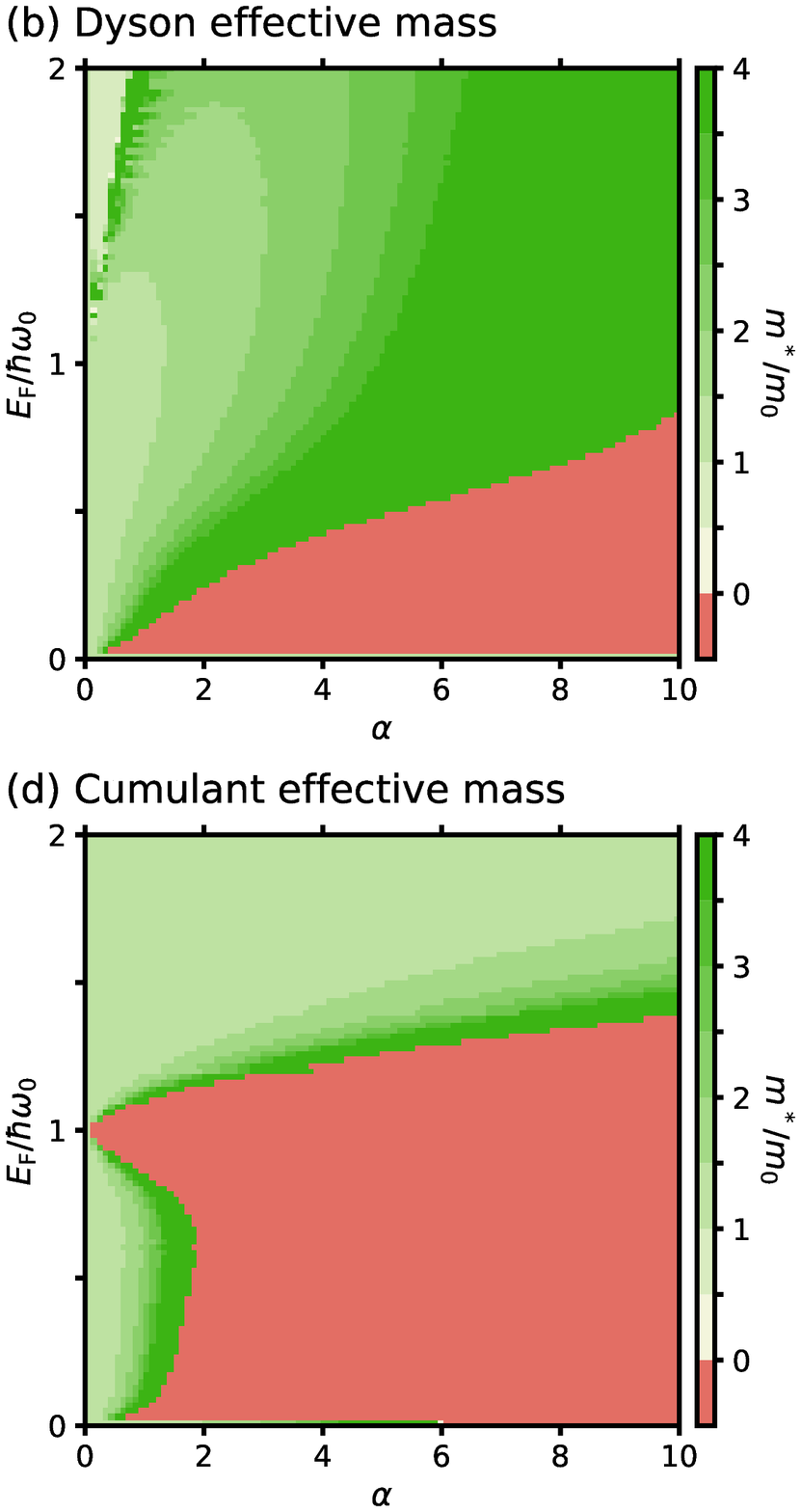}
  \end{minipage}
  \caption{
  (a) Renormalized Dyson quasi-particle energy $E_0/\hbar\omega_0$
  relative to the Fermi level, including screening by free carriers. Negative
  QP energies (shown in green) indicate a higher binding
  energy of the interacting system, whereas positive energies
  (in red) imply 
  that the renormalized QP energy lies 
  above the Fermi level, and indicate a breakdown of the Fermi surface.
  (b) Renormalized second-order cumulant quasi-particle energy relative to
  the Fermi surface.
  (c) Renormalized Dyson effective mass at the $\Gamma$ point.
  Positive values of $m^*$ (shown in green) indicates the renormalization of the 
  QP mass; negative values of $m^*$ (shown in red) indicate that the curvature
  of the QP spectrum at $k=0$ has become negative, and signal a breakdown
  of the second-order expansion of the self-energy. 
  (d) Renormalized second-order cumulant effective mass.}
  \label{fig9}
  \end{figure*}

\end{document}